\documentclass[hyper]{JHEP3}

\usepackage{latexsym,amsmath,amsfonts,amssymb}
\usepackage{graphicx}
\usepackage[all]{xy}

\newcommand{\ds}{\displaystyle}
\newcommand{\ts}{\textstyle}
\newcommand\eqr[1]{(\ref{#1})}

\newcommand{\del}{\partial}

\newcommand{\be}{\begin{equation}}
\newcommand{\ee}{\end{equation}}
\newcommand{\beq}{\begin{equation}}
\newcommand{\eeq}{\end{equation}}
\newcommand{\bea}{\begin{eqnarray}}
\newcommand{\eea}{\end{eqnarray}}

\newcommand{\al}{\alpha}

\newcommand{\cC}{\mathcal{C}}

\newcommand{\cF}{\mathcal{F}}

\newcommand{\cL}{\mathcal{L}}
\newcommand{\cM}{\mathcal{M}}
\newcommand{\cN}{\mathcal{N}}
\newcommand{\cO}{\mathcal{O}}

\newcommand{\cS}{\mathcal{S}}

\newcommand{\bC}{\mathbb{C}}
\newcommand{\bH}{\mathbb{H}}

\newcommand{\bP}{\mathbb{P}}
\newcommand{\bR}{\mathbb{R}}
\newcommand{\bZ}{\mathbb{Z}}

\newcommand\cf{{\it c.f.}}
\newcommand\ie{{\it i.e.}}
\newcommand\eg{{\it e.g.}}

\DeclareMathOperator{\Tr}{Tr}

\title{Four-Dimensional SCFTs from M5-Branes}
\author{Ibrahima Bah$^1$, Christopher Beem$^{2}$,  Nikolay Bobev$^{2}$, Brian Wecht$^{3,4}$ \\

{${}^1$Michigan Center for Theoretical Physics} \\ 
{\,\;University of Michigan} \\
{\,\;Ann Arbor, MI 48109 USA} \\
{$\,$} \\
{${}^2$Simons Center for Geometry and Physics}\\
{\,\;Stony Brook University}\\
{\,\;Stony Brook, NY 11794-3636, USA}\\
{$\,$}\\
{${}^3$Center for the Fundamental Laws of Nature}\\
{\,\;Harvard University}\\
{\,\;Cambridge, MA 02138, USA}\\
{$\,$}\\
{${}^4$Centre for Research in String Theory}\\
{\,\;Queen Mary, University of London}\\
{\,\;London E1 4NS, UK}\\
{$\,$}}

\abstract{We engineer a large new set of four-dimensional $\cN=1$ superconformal field theories by wrapping M5-branes on complex curves. We present new supersymmetric $AdS_5$ M-theory  backgrounds which describe these fixed points at large $N$, and then directly construct the dual four-dimensional CFTs for a certain subset of these solutions. Additionally, we provide a direct check of the central charges of these theories by using the M5-brane anomaly polynomial. This is a companion paper which elaborates upon results reported in \cite{Bah:2011vv}.}

\preprint{MCTP-12-05, QMUL-PH-12-05}
\begin{document}
\setcounter{page}{0}
\section{Introduction}
\label{sec:intro}

The M5-brane is one of the most interesting and enigmatic denizens of the string/M-theory menagerie. Although we know some elementary properties of the six-dimensional $(2,0)$-supersymmetric theory living on the worldvolume of $N$ coincident M5-branes -- \eg, that the number of degrees of freedom scales as $N^3$ -- many aspects of this theory remain poorly understood. One strategy for obtaining a window into such theories is to wrap the branes on a complex curve and study the resulting four-dimensional infrared effective theory. This compactification can generically be done so as to preserve as many as eight supercharges, in which case one can hope to elucidate the resulting theories by leveraging the many powerful tools that have been developed for studying four-dimensional $\cN=2$ gauge theories. This approach has been taken first in \cite{Gaiotto:2009we,Gaiotto:2009hg} and then in \cite{Alday:2009qq,Kanno:2009ga,Gaiotto:2010be,Tachikawa:2010vg,Nanopoulos:2010ga,Chacaltana:2010ks,Gaiotto:2011tf,Cecotti:2011rv,Tachikawa:2011yr,Alim:2011ae,Gaiotto:2011xs,Alim:2011kw}, often with remarkable efficacy.

It is also possible to wrap M5-branes on a complex curve and preserve only four supercharges; the IR theory may then potentially realize a four-dimensional $\cN=1$ superconformal field theory. Because of the reduced amount of supersymmetry, these theories are less accessible than their $\cN=2$ counterparts. Nevertheless, it is possible to make a great deal of progress. The foundational work of Maldacena and N\'u\~nez (MN) \cite{Maldacena:2000mw} exhibited solutions of eleven-dimensional supergravity preserving either $\cN=1$ or $\cN=2$ supersymmetry which are holographically dual to the theory of M5-branes wrapped on a closed genus $g$ curve $\cC_g$. In the near-horizon limit, these solutions have the form $AdS_5 \times \cC_g \times  S^4$, where $S^4$ is a squashed four-sphere that is fibered over $\cC_g$. Although the $AdS_5$ region indicates the existence of an interacting conformal phase in the IR limit of the M5-brane theories, the CFTs remained largely mysterious for many years.

Almost a decade later, Gaiotto provided the crucial missing piece essential to understanding these four-dimensional SCFTs \cite{Gaiotto:2009we}. He gave evidence for the existence of a family of isolated $\cN=2$ SCFTs, denoted $T_N$, which arise on the worldvolume of $N$ coincident M5-branes wrapping a thrice-punctured sphere. These theories possess $SU(N)^3 \times SU(2)_R \times U(1)_R$ global symmetry, and have no known weakly coupled description. In the follow-up work \cite{Gaiotto:2009gz}, it was argued that the SCFT duals to the $\cN=2$ MN solutions could be constructed by combining $T_N$ theories via gauging diagonal subgroups of various $SU(N) \times SU(N)$ global symmetries. This diagonal gauging corresponds to sewing together the punctured spheres to form a more complicated Riemann surface.

Although the $T_N$ SCFT is an $\cN=2$ theory, one can use it to construct and study various $\cN=1$ theories as long as supersymmetry is broken in a controlled way. For example, if the $\cN=2$ theory is deformed by an $\cN=1$ superpotential, then the global symmetries of the IR $\cN=1$ theory are known, assuming that there are no accidental symmetries. This was the philosophy adopted in \cite{Benini:2009mz}, where it was argued that the SCFT dual to the $\cN=1$ MN solution could be obtained by adding a mass term for the $\cN=1$ adjoint chiral superfields in the $\cN=2$ vector multiplets of \cite{Gaiotto:2009gz}. The result is a theory where $T_N$ blocks are connected together using only $\cN=1$ vector multiplets, and there is a nonvanishing superpotential. Additionally, in \cite{Bah:2011je}, a large number of new $\cN=1$ theories involving $T_N$'s were constructed, where $T_N$ building blocks are connected by both vector and hypermultiplets. 

In this paper, we present and study an infinite class of four-dimensional $\cN=1$ SCFTs associated to a complex curve $\cC_g$ that are a natural generalization of the MN theories and, in a precise sense which we describe later, interpolate between and then extend beyond the MN solutions. There are three complementary vantage points from which to view the SCFTs in question, all of which are useful for establishing the existence of the theories and ascertaining some of their properties.

The first of these viewpoints, discussed in Section \ref{sec:six}, is to consider coincident M5-branes wrapping a holomorphic curve. From this perspective, the generalization of the MN theories arises from the wide variety of possible arrangements of the normal geometry of a given curve in a Calabi-Yau threefold. In particular, for any local geometry of the form $\cL_1\oplus\cL_2\rightarrow\cC_g$ for line bundles $\cL_1$ and $\cL_2$ of degrees $p$ and $q$ (with $p+q=2g-2$), there is an associated $\cN=1$ SCFT.\footnote{The cases $g=0,1$ are special. For $g=0$, there is a nontrivial fixed point only if $|p-q|\geq4$. For $g=1$, the case of $p=q=0$ simply leads to $\cN=4$ SYM, while all other arrangements lead to new fixed points.} We compute the central charges of these theories at finite $N$ using the anomaly polynomial of the M5-brane. In an interesting extension of the analogous calculation in \cite{Benini:2009mz}, the presence of an additional global $U(1)$ symmetry in each such construction introduces some uncertainty into the identification of the  superconformal R-symmetry, necessitating the use of $a$-maximization \cite{Intriligator:2003jj}.

The second perspective, discussed in Sections \ref{sec:seven} and \ref{sec:eleven}, is the holographic one. The internal space of the MN solutions is a squashed $S^4$ bundle over the curve $\cC_g$, where the $S^4$ is squashed in such a way as to preserve (at least) $U(1)^2$ isometry. The generalizations we describe are of the same form, where the $S^4$ fibration and squashing are modified but still respect a $U(1)^2$ symmetry. In contrast with the MN solutions, our more general backgrounds include the case where $\cC_g$ is a two-sphere or a flat torus. The $S^4$ fibration reflects the structure of the Calabi-Yau geometry in which the M5-branes are embedded before backreaction. The large $N$ perspective also allows us to establish that the fixed points in question are realized dynamically by studying holographic RG flows which interpolate between a UV region describing the six-dimensional $(2,0)$ theory and the fixed point solution in the IR. By recasting the solutions in the canonical form of \cite{Gauntlett:2004zh}, we can further observe that the superconformal R-symmetry of these backgrounds matches that produced by $a$-maximization. 

The final approach is a purely four-dimensional one in the spirit of the $\cN=1$ generalized quiver constructions of \cite{Bah:2011je}. In Section \ref{sec:four}, we find that when $T_N$ building blocks are combined with both $\cN=1$ and $\cN=2$ vector multiplets in just the right fashion, the resulting theory preserves one additional $U(1)$ global symmetry. Moreover, the central charges of the infrared fixed point exactly match the ones computed from the M5-brane anomaly polynomial when $g>1$ and $p$ and $q$ are both non-negative. We also study the spectrum of relevant and marginal operators and compute the dimension of the superconformal manifold of these theories. The number of independent exactly marginal operators matches the expectation from the gravity side, although we find a puzzle having to do with the counting of relevant operators. Finally, in Section \ref{sec:example}, we work through some examples for the $g=3$ quivers in detail.

All of the field theory tools we use rely only on global symmetries, and we do not need a weakly coupled UV Lagrangian description. In particular, because the generalized quivers we study make use of the $T_N$ theory, they fall into the category of ``non-Lagrangian'' SCFTs.  Although we do not have a field theory construction of the CFT duals for all of our $AdS_5$ solutions, we expect that the absence of a weakly coupled description should be a generic property of all such models; this is the na\"ive expectation for any such theory arising from a nontrivial compactification of the six-dimensional $(2,0)$ theory. Thus, the theories presented here dramatically increase the number of known ``non-Lagrangian'' $\cN=1$ theories, thereby providing more evidence that such theories may make up a sizable portion of the landscape of four-dimensional SCFTs.

\section{M5-branes on complex curves}
\label{sec:six}

The twisted versions of supersymmetric field theories have been a subject of great interest and a source of many insights in physics and mathematics since their introduction \cite{Witten:1988ze}. Many such constructions arise naturally in string theory through the low-energy dynamics of branes wrapping curved, supersymmetric cycles in manifolds with special holonomy \cite{Bershadsky:1995qy}. In the language of wrapped branes, the details of the twist are encoded in the normal geometry to the wrapped cycle. In particular, the conditions for unbroken supersymmetry impose relations between the curvature of the normal bundle and that of the tangent bundle, such that they allow for the existence of covariantly constant spinors on the brane worldvolume (or alternatively, render the neighborhood of the brane Ricci-flat). In this section, we describe the class of twisted field theories we will be studying in terms of their geometric brane realization in M-theory. We further explain that under certain assumptions, one can compute the central charges of the resulting low-energy fixed points.

\subsection{The local geometry}\label{localgeometry}

In many situations, the normal geometry to a calibrated cycle is completely fixed in terms of the geometry of the cycle \cite{Gauntlett:2003di}. For instance, two popular examples of wrapped brane constructions are M5-branes wrapping holomorphic curves in a Calabi-Yau twofold (\cf\ \cite{Gaiotto:2009we}) or special Lagrangian three-cycles in a Calabi-Yau threefold (\cf\ \cite{Dimofte:2011ju}). In these geometries, the normal bundle is restricted to be precisely the canonical or cotangent bundle to the cycle, respectively, and the only freedom present in the choice of local geometry is the choice of metric on the cycle itself. 

Alternatively, in the case of a genus $g$ holomorphic curve $\cC_g$ in a Calabi-Yau threefold $X$, the local geometry takes the form of a holomorphic $\bC^2$ bundle over the curve,
\be\label{local} 
\xymatrix{
\bC^2 \ar[r] & X \ar[d]\\
& \cC_g\\}
\ee
and the requirement for supersymmetry to be preserved is that the determinant line bundle of $X$ be the canonical bundle $K_{\cC_g}$ of the curve. This means that in terms of the $U(2)$-valued connection on the vector bundle $X$, only the central $U(1)$ is constrained. Consequently, $X$ takes the form of the following tensor product of vector bundles,
\be\label{localprod}
X=K_{\cC_g}\otimes V~,
\ee
where $V$ is an {\it arbitrary} $SU(2)$-bundle over the curve. The goal of this section is to explore how the infrared dynamics on the wrapped M5-branes depends on the choice of this bundle.

A simple example is sufficient to demonstrate that different choices for $V$ can lead to inequivalent four-dimensional theories in the infrared. When $V$ is a {\it flat} $SU(2)$-bundle, the dynamics reduces at low energies to the family of $\cN=1$ SCFTs studied in \cite{Benini:2009mz}. On the other hand, by choosing an appropriately curved $SU(2)$ bundle, the total space can be made to equal the direct (Whitney) sum of a trivial line bundle with the canonical bundle. In this case, the threefold geometry simplifies to a direct product of the complex plane with the cotangent bundle of $\cC_g$, 
\be\label{N=2geom}
X=\bC\times T^{\star}\cC_g~.
\ee
In this case, twice as much supersymmetry is preserved and the low energy theory is an $\cN=2$ SCFT ``of class $\cS$'' \cite{Gaiotto:2009we,Gaiotto:2009hg}.

The question is then whether there are still more possibilities which lead to new infrared fixed points. It is not immediately obvious that this should be the case, as data describing the local geometry of the curve in six dimensions can be irrelevant under the renormalization group flow to four dimensions. In particular, aside from the choice of complex structure, the metric on $\cC_g$ has been proven to be irrelevant in certain cases \cite{Anderson:2011cz}. Nevertheless, as we now argue, there is good reason to believe that for a given curve $\cC_g$ there exist infinitely many inequivalent fixed points which can be obtained from six dimensions for appropriate choices of $V$. 

The primary observation is that if the structure group of $V$ reduces from $SU(2)$ to $U(1)$, then $X$ is {\it decomposable}, and the local geometry \eqr{local} enjoys an additional Abelian symmetry $U(1)_\cF$ under which the preserved supercharges of the M5-brane theory are invariant. This is in addition to the omnipresent R-symmetry $U(1)_{K}$ which acts as phase rotations of $K_{\cC_g}$ in \eqr{localprod}. More specifically, the space $X$ will take the simple form
\be\label{locallines} 
\xymatrix{
\bC^2 \ar[r] & \cL_1\oplus\cL_2 \ar[d]\\
& \cC_g\\}\ee
subject to the condition that $\cL_1\otimes\cL_2 = K_{\cC_g}$. There is a manifest $U(1)_1\times U(1)_2$ symmetry, where $U(1)_i$ act as phase rotations on the fibers of the line bundle $\cL_i$. In terms of these symmetries, the $U(1)_K$ and $U(1)_\cF$ act as the symmetric and antisymmetric combinations of $U(1)_1$ and $U(1)_2$, respectively.

There are an infinite number of disjoint families of such decomposable bundles, labeled by a choice of integer Chern numbers, 
\be\label{chernnumbers}
c_1(\cL_1)=p~,\qquad\qquad c_1(\cL_2)=q~,
\ee
satisfying $p + q = 2g-2$. For different choices of $p$ and $q$, the fields of the M5-brane theory transform in different representations of $U(1)_\cF$. It is thus natural to expect that the corresponding four-dimensional infrared fixed points will be distinct. This expectation is clarified and confirmed below when we discuss the anomalies and central charges of these theories.

\subsection{Central charges from six dimensions}\label{subsec:cc6}

Because so little is known about the six-dimensional $(2,0)$ theory, not many direct calculations can be done to study the low energy physics of the configurations described above. Nevertheless, it is possible to compute the central charges of the infrared fixed points given the assumption that there are no accidental symmetries which appear along the renormalization group flow. The calculation which we describe proceeds along the lines of \cite{Alday:2009qq,Benini:2009mz}, with the additional complication that because of the extra $U(1)_\cF$ symmetry, $a$-maximization must be utilized to determine the appropriate superconformal R-symmetry \cite{Intriligator:2003jj}. 

In four-dimensional $\cN=1$ SCFTs, the central charges $a$ and $c$ are completely determined by the linear and cubic 't Hooft anomalies of the superconformal R-symmetry \cite{Anselmi:1997am},
\be\label{acdef}
a = \tfrac{3}{32} \left(3  \Tr R^3 -  \Tr R \right)~,\qquad\qquad c = \tfrac{1}{32} \left(9  \Tr R^3 - 5 \Tr R \right)~.
\ee
These anomalies are conveniently packaged in terms of an anomaly six-form -- related to the anomalous divergence of the R-current by the descent procedure -- which reads
\be\label{sixform}
I_6=\frac{\Tr R^3}{6}c_1(F)^3-\frac{\Tr R}{24}c_1(F)p_1(T_4)~,
\ee
where $F$ is the $S^1$ bundle which couples to the superconformal R-symmetry, and $p_1(T_4)$ is the first Pontryagin class of the tangent bundle to the four-dimensional spacetime manifold on which the theory is defined. 

The key fact which allows us to compute the four-dimensional central charges of our theories is that this anomaly six-form can be obtained by starting with the anomaly eight-form of the M5-brane theory and integrating over $\cC_g$. Because the eight-form encodes the anomalous  diffeomorphisms of the M5-brane worldvolume and its normal bundle, this approach will work as long as the four-dimensional R-symmetry appearing in \eqr{sixform} is realized geometrically in the brane setup. In the constructions described above, there is an additional subtlety due to the presence of the global symmetry $U(1)_\cF$. The R-symmetry appearing in \eqr{sixform} will then be of the form
\be\label{Rtrial}
R=K+\epsilon \cF~,
\ee
where $\epsilon$ is a real number determined by $a$-maximization \cite{Intriligator:2003jj}. As we will see, the central charges have a nontrivial dependence on the choice of local geometry \eqr{locallines}, confirming our expectations.

One might wonder whether for $g=0$ and $g=1$, symmetries of the curve $\cC_g$ lead to additional Abelian symmetries in the field theory that could potentially mix with the R-symmetry. For $g=0$, this cannot be the case as long as the full $SU(2)$ isometry group of $S^2$ is preserved, since a non-Abelian symmetry will not mix with the R-symmetry. Indeed, the gravity solutions of Section \ref{sec:seven} demonstrate that the the full $SU(2)$ isometry of the sphere is preserved. For $g=1$ and $z\neq0$, the symmetry group will necessarily be Abelian, but any mixing of these symmetries with the R-symmetry would not be invariant under the $SL(2,\bZ)$ modular group of the torus.\footnote{The case $g=1$ and $z=0$ is special, and leads to $\cN=4$ SYM in the infrared.} 

With this caveat out of the way, we return to the calculation of the anomaly for M5-branes on $\cC_g$. The anomaly eight-form for a single M5-brane is \cite{Witten:1996hc}
\be
I_{8}[1] = \ds\tfrac{1}{48} \left[ p_2(NW) - p_2(TW) +\ds\tfrac{1}{4}(p_1(TW) - p_1(NW) )^2 \right]~,
\ee
where $p_k$ is the $k$-th Pontryagin class, $TW$ denotes the tangent bundle to the worldvolume of the M5-brane, and $NW$ is the normal bundle. For a $(2,0)$ theory of type $G=A_N,D_N,E_{6,7,8}$, the anomaly takes the form \cite{Harvey:1998bx,Yi:2001bz,Intriligator:2000eq}
\be\label{eightform}
I_{8}[G] = r_G I_{8}(1) + \ds\frac{d_Gh_G}{24} p_2(NW)~,
\ee
where now the normal bundle $NW$ can be thought of as an $SO(5)$ bundle coupled to the R-symmetry of the six-dimensional theory. Here, $r_G,~d_G,~h_G$ are the rank, dimension, and Coxeter number of $G$, respectively (see Table \ref{table1}).


\begin{table}[ht]
\centering
\begin{tabular}{|c|c|c|c|}
\hline
$G$ & $r_G$ & $d_G$ & $h_G$ \\[0.2ex]
\hline
$A_{N-1}$ & $N-1$ & $N^2-1$ & $N$ \\
$D_{N}$ & $N$ & $N(2N-1)$ & $2N-2$ \\
$E_6$ & $6$ & $78$ & $12$ \\
$E_7$ & $7$ & $133$ & $18$ \\
$E_8$ & $8$ & $248$ & $30$ \\ [0.2ex]
\hline
\end{tabular}
\caption{Simply Laced Lie Algebras}
\label{table1}
\end{table}


The first and second Pontryagin classes of a vector bundle $E$ are expressed in terms of the Chern roots $e_i$ as
\be
p_1(E) = \ds\sum_{i}e_i^2~, \qquad\qquad p_2(E) = \ds\sum_{i<j}e_i^2e_j^2~.
\ee
For the geometries of interest, the Chern roots of the normal bundle are simply those of the line bundles $\cL_1$ and $\cL_2$ in \eqr{locallines}, which have Chern numbers $p$ and $q$, respectively. The Calabi-Yau condition then imposes $p + q = 2g-2$, the solutions to which we parameterize as
\be\label{pqdef}
p=(1+z)(g-1)~,\qquad\qquad q=(1-z)(g-1)~,
\ee
with 
\be\label{zdef}
z=\frac{n}{(g-1)}~,\qquad n\in\bZ~.
\ee
We denote the Chern roots of the tangent bundle by $\lambda_1$, $\lambda_2$, and $t$, where $t$ is the tangent bundle to $\cC_g$ and $\lambda_{1,2}$ are the Chern roots of the four-dimensional tangent bundle $T_4$ in equation \eqr{sixform}.

To compute the anomaly six-form for a $U(1)$ symmetry of the form \eqr{Rtrial}, we couple the symmetry to a nontrivial $U(1)$ bundle $F$ over the flat four-dimensional part of the brane worldvolume. This introduces a shift in the Chern roots,
\be
\cL_1 \to \cL_1 + (1+\epsilon) F~, \qquad\qquad  \cL_2 \to \cL_2 + (1-\epsilon) F~.
\ee
Here $\epsilon$ is a real parameter which we keep general for the moment since we do not know the correct value for which \eqr{Rtrial} becomes the superconformal R-symmetry.

The eight-form \eqr{eightform} can then be integrated over $\cC_g$ to obtain the effective anomaly six-form \eqr{sixform},\footnote{Here we use the fact that $\int_{\cC_g} c_1(t) = 2-2g$, and also assume that $g\neq1$. The calculation for $g=1$ follows the same lines, but leads to somewhat different results. We discuss this case separately in Appendix \ref{app:T2}.}
\begin{align}\begin{split}
\int_{\cC_g} I_{8} = \ds\frac{g-1}{6} [ (r_G+d_Gh_G) (1+ z \epsilon^3) - &d_G h_G (\epsilon^2 + z \epsilon) ] c_1(F)^3\\ 
&- \ds\frac{g-1}{24} r_G (1+ z \epsilon) c_1(F) p_1(T_4)~.
\end{split}\end{align}
From this expression it is easy to read off the R-anomaly of the four-dimensional theory,
\be
\Tr R^3 =  (g-1) [ (r_G+d_Gh_G) (1+ z \epsilon^3) - d_G h_G (\epsilon^2 + z \epsilon) ]~, \quad \Tr R =  (g-1)r_G (1+z \epsilon) ~.
\ee
Now we can use the expressions in \eqr{acdef} to compute trial central charges as functions of the parameter $\epsilon$. The correct value of $\epsilon$ is then determined by maximizing $a$, yielding%
\be
\epsilon = \ds\frac{\eta +\kappa \zeta}{3 (1+\eta) z}~,
\label{maxepsilon}
\ee
where we have defined the parameter $\kappa=1$ for $S^2$ and $\kappa=-1$ for a hyperbolic Riemann surface. We have also defined
\be
\eta \equiv h_G(1+h_G)~, \qquad\qquad \zeta \equiv \sqrt{\eta^2 +(1+4\eta+3\eta^2)z^2}~,
\label{etazetadef}
\ee
and used the group theory identity
\be
d_G = r_G (h_G+1)~.
\ee
The result for the central charges is
\begin{align}\begin{split}\label{exactcentral}
a &= (g-1) r_G \ds\frac{\zeta^3+\kappa \eta^3  -\kappa(1+\eta)(9+21 \eta + 9 \eta^2) z^2  }{48 (1+\eta)^2 z^2}~,\\
c &= (g-1) r_G \ds\frac{\zeta^3+\kappa \eta^3 -\kappa(1+\eta)(6-\kappa \zeta+17 \eta + 9 \eta^2) z^2}{48 (1+\eta)^2 z^2}~.
\end{split}\end{align}
For the $A_N$ theory in the large $N$ limit, the parameter $\epsilon$ is
\be
\epsilon = \ds\frac{1+\kappa \sqrt{1+3z^2}}{3z}~,
\label{epslargeN}
\ee
and the central charges simplify to
\be
a=c = {(1-g)N^3}\left(\ds\frac{1- 9z^2+\kappa(1+3\, z^2)^{3/2}}{48 \,z^2}\right)~. 
\label{aclargeNanomaly}
\ee

\begin{figure}[t]
\centering
\includegraphics[width=10cm]{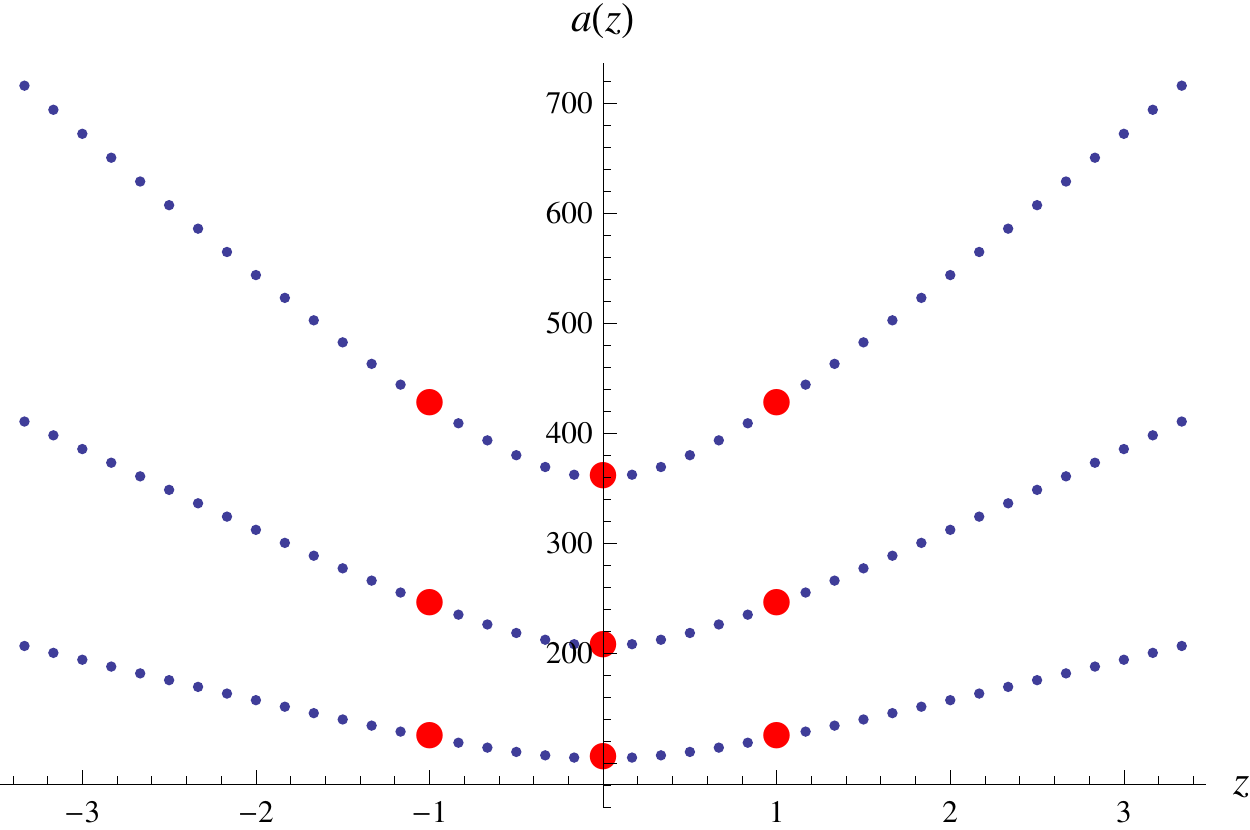}
\caption{\it The central charge a as a function of the twist parameter $z$ for genus $g=7$, $G=A_{N-1}$, and $N=4,5,6$ (bottom to top). The MN theories are marked with large points at $z=0$ and $|z|=1$.}
\label{ccplotanom}
\end{figure}

For $|z|= 1$, the Calabi-Yau geometry is that of \eqr{N=2geom} -- corresponding to the $\cN=2$ MN theory -- and the central charges match those computed in \cite{Gaiotto:2009gz}. For $z=0$, the geometry is given by \eqr{localprod} with $V$ a flat (in fact, trivial) bundle, and the central charges match those of the $\cN=1$ MN theory. More generally, for any choice of $g$ and $z$ these results suggest the existence of an interacting SCFT with central charges given by \eqr{exactcentral}. For fixed $g>1$, we find $g-2$ distinct fixed points ($0<z<1$) with central charges taking intermediate values between those of the $\cN=1$ and $\cN=2$ MN fixed points. We further find an infinite number of fixed points for $z>1$ with central charges greater than those of the $\cN=2$ MN fixed points, obeying $a\sim c\sim z$ for $z\gg1$ (see Figure \ref{ccplotanom}).\footnote{It is interesting to observe that by Riemann-Roch, for large $z$ the number of holomorphic deformations of the curve $\cC_g$ in $X$ grows linearly in $z$. This suggests that the growth of central charge can be (at least partially) accounted for by an increasing number of light degrees of freedom describing these deformations.}

For $g=0$, the parameter $z$ is an integer and for $z=0,\pm1$ the central charges in \eqr{exactcentral} are negative, indicating the presence of accidental symmetries which render our $a$-maximization procedure incorrect. This is in harmony with the na\"ive expectation that the low energy field theories on branes wrapping positively curved manifolds should confine. However, we find that for $|z|>1$, the computed central charges do not violate unitarity, potentially signaling the existence of an interacting conformal phase for M5-branes wrapping a rational curve in a Calabi-Yau threefold with an unstable normal bundle of the form $\cO(-3-n)\oplus\cO(1+n)$ with $n\geq0$. For precisely these choices of normal bundle, the zero-section is not contractible to a point, so again  this result fits with the intuition that confinement arises from the wrapped cycle ``shrinking'' and being replaced via a geometric transition.\footnote{We thank Dave Morrison for pointing out to us that this does indeed hold true for the choice $n=0$.}

There is one additional wrinkle in this story. Along with the positivity of $a$ and $c$, which is manifest from \eqref{exactcentral}, there are several bounds which are known to constrain the possible values of the central charges of four-dimensional CFTs. In particular, for an $\cN=1$ SCFT it was argued in \cite{Hofman:2008ar} (and further supported in \cite{Kulaxizi:2010jt}) that they must obey
\be\label{acbound}
\ds\frac{1}{2}\leq\ds\frac{a}{c}\leq\ds\frac{3}{2}~.
\ee
Interestingly, for all but one choice of parameters $(g,z,N)$ with positive $a$ and $c$, this bound is satisfied. However, for $g=0$, $z=2$, and $N=2$, the bound is violated and we find $a/c\approx .387641$. We do not have an interpretation of what goes wrong in this special case, and it would be of great interest to learn whether or not whatever issues plague this theory are also a problem for other choices of parameters. For good measure, we note that all central charges in \eqref{exactcentral} obey the bound on $c$ studied in \cite{Rattazzi:2010gj,Poland:2011ey}. 

\section{Holographic duals from gauged supergravity}
\label{sec:seven}

In the case of the MN twists, the gravity dual description predated any direct field theoretic analysis, offering insights into the nature of the four-dimensional fixed points at large $N$ \cite{Maldacena:2000mw}. We will similarly be able to find dual supergravity backgrounds which describe the backreaction of M5-branes in the geometries discussed above.
This establishes at large $N$ the existence of the superconformal fixed points discussed in the previous section. Moreover, there are holographic RG flows which demonstrate that the fixed point theories are realized dynamically. For concreteness we concentrate on the $A_{N}$ $(2,0)$ theory, but our discussion could be easily adapted to accommodate the $D_N$ theory as well. We briefly comment on this in Section \ref{sec:eleven}.

The six-dimensional $(2,0)$ theory of type $A_N$  is dual at large $N$ to eleven-dimensional supergravity on $AdS_7\times S^4$. The large $N$ dual of a compactification of this theory on $\cC_g$ is then eleven-dimensional supergravity in a background which is asymptotically locally $AdS_7\times S^4$, but for which the topology at fixed value of the radial coordinate is an $S^4$ fibration over $\mathbb{R}^{1,3}\times\mathcal{C}_g$ \cite{Maldacena:2000mw}. The $S^4$ fibration at the boundary is precisely specified by the $\mathbb{R}^5$ fibration in the brane construction, in that they are determined by the same $SO(5)$ connection on $\cC_g$. Eleven-dimensional supergravity on $S^4$ admits a consistent truncation to the lowest Kaluza-Klein modes given by the maximal gauged supergravity in seven dimensions \cite{Nastase:1999cb,Nastase:1999kf}. The operators that are turned on in the M5-brane theory upon compactification are dual to a subset of these modes. Therefore the existence of the consistent truncation guarantees that the gravity duals to the constructions of the previous section are captured by the lower-dimensional gauged supergravity. Moreover, all of the solutions we obtain can be uplifted to solutions in eleven dimensions using explicit formulae from \cite{Nastase:1999cb,Nastase:1999kf,Cvetic:1999xp}.

The supergravity fields which are needed to describe the partial twists in question lie in a further truncation of the maximal seven-dimensional supergravity with bosonic fields consisting of the metric, two Abelian gauge fields in the Cartan of the $SO(5)$ gauge group, and two real scalars which parameterize squashing deformations of the $S^4$. The Abelian gauge symmetries of this truncation correspond directly to the global symmetries $U(1)_1$ and $U(1)_2$ of Section \ref{localgeometry}. This supergravity truncation was studied in \cite{Liu:1999ai}, and it was established in \cite{Cvetic:1999xp,Liu:1999ai} that every solution of the equations of motion of the truncated theory solves the equations of motion of the maximal theory.

Imposing the most general Ansatz that will capture the solutions of interest, the seven-dimensional metric takes the form
\be
ds^2 = e^{2f} (-dt^2+dz_1^2+dz_2^2+dz_3^2) + e^{2h}dr^2 + e^{2\hat{g}} (dx_1^2+dx_2^2)~, \label{M5metricAnsatz}
\ee
where $f$, $\hat{g}$, and $h$ are functions of the radial coordinate $r$ and of the coordinates $(x_1,x_2)$ on the Riemann surface. In general there are also nontrivial $(r,x_1,x_2)$-dependent profiles for the two Abelian gauge fields and the two real scalars,
\be
 A^{(i)}= A^{(i)}_{x_1} dx_1 +A^{(i)}_{x_2} dx_2+A^{(i)}_r dr~, \qquad\qquad \lambda_i=\lambda_i(x_1,x_2,r)~, ~~i=1,2~. \label{M5gaugescalarAnsatz}
\ee
The asymptotic form of this Ansatz is determined by the brane construction of the dual field theory. In particular, the metric functions $f$, $\hat{g}$, and $h$ should be asymptotic to $\log(1/ r)$ to ensure that the space is asymptotically locally $AdS_7$ in the UV. The two real scalars should vanish and the Abelian gauge fields are equal to the connections on $\cL_1$ and $\cL_2$ in \eqr{locallines}. In the IR we find supersymmetric $AdS_5$ solutions for each choice of the parameter $z$ in \eqr{zdef}. In the remainder of this section, we describe these fixed point solutions as well as the supergravity domain walls which interpolate between them and the twisted theory in the UV.

\subsection{$\mathcal{N}=1$ fixed points}
\label{subsec:N=1fp}

The BPS equations which govern supersymmetric solutions of the form \eqr{M5metricAnsatz}-\eqr{M5gaugescalarAnsatz} are derived in Appendix \ref{app:BPSeqns}. With the additional assumption that \eqr{M5metricAnsatz} describes a warped product $AdS_5\times\cC_g$, we find that $h(x_1,x_2,r)=f(x_1,x_2,r)=-\log r+f_0$. Furthermore, the metric on $\cC_g$ is restricted to be of constant curvature $\kappa=\pm1,0$,\footnote{We will primarily discuss the cases $\kappa=\pm1$ in this text. The case of $\kappa=0$ (\ie\ $\cC_g$ a flat torus) is discussed in Appendix \ref{app:T2}.}
\bea
\hat{g}(x_1,x_2,r) = &g_0 - \log x_2\;\;\qquad\qquad\qquad~&, \qquad \kappa=-1~, \\
\hat{g}(x_1,x_2,r) = &g_0 - 2\log\left(\ds\frac{1+x_1^2+x_2^2}{2}\right)~&, \qquad \kappa=+1 ~. 
\eea
From now on for $\kappa=-1$, where $(x_1,x_2)$ parameterize the upper half plane, we quotient $\mathbb{H}^2$ by an appropriate Fuchsian subgroup  $\Gamma\in PSL(2,\mathbb{R})$ resulting in a closed Riemann surface with genus $g>1$. The gauge fields have field strengths given by
\bea
F^{(1)}_{x_1x_2} &=& \ds\frac{1}{8g-8}\ds\frac{p}{x_2^2}~, \qquad\qquad F^{(2)}_{x_1x_2}= \ds\frac{1}{8g-8}\ds\frac{q}{x_2^2}~, \qquad \kappa=-1~, \\
F^{(1)}_{x_1x_2} &=& \ds\frac{1}{2}\ds\frac{p}{(1+x_1^2+x_2^2)^2}~, \quad F^{(2)}_{x_1x_2}= \ds\frac{1}{2}\ds\frac{q}{(1+x_1^2+x_2^2)^2}~, \quad  \kappa=1~,
\eea
where $p$ and $q$ are as in \eqr{chernnumbers}. Finally, the real scalar fields take constant values
\be
\lambda_1=\lambda_{1}^{(0)}~, \qquad\qquad \lambda_2=\lambda_{2}^{(0)}~.
\ee
The most general $AdS_5$ solution of the BPS equations is then parameterized by the genus $g$ and the ``twist parameter'' $z$, and is given by
\begin{align}
\begin{split}
e^{10\lambda_1^{(0)}} &= \ds\frac{1+7\,z+7 \,z^2 + 33\,z^3 + \kappa(1+4\,z+19\,z^2)\sqrt{1+3\,z^2}}{4\,z(1-z)^2}~,\\
e^{2(\lambda_1^{(0)}-\lambda_2^{(0)})} &= \ds\frac{1+z}{2\,z-\kappa \sqrt{1+3\,z^2}}~,\\
e^{2g_0} &= - \ds\frac{\kappa}{8}\, e^{2\lambda_1^{(0)}+2\lambda_2^{(0)}} \left(\left(1-z\right)e^{2\lambda_1^{(0)}} + \left(1+z\right) e^{2\lambda_2^{(0)}}\right)~,\\
e^{f_0} &= e^{4(\lambda_1^{(0)}+\lambda_2^{(0)})}~. 
\label{N=1fixedpoints}
\end{split}
\end{align}
The twist parameter $z$ is quantized as in \eqr{zdef} so as to ensure consistency of the gauge fields on $\cC_g$.

For $\kappa=-1$, the metric functions and the scalars are real for all properly quantized values of $z$. If $\kappa=1$, the metric functions are real and finite only for $|z| > 1$, \ie, there are $AdS_5\times S^2$ $\mathcal{N}=1$ vacua for $|z| > 1$. It is worth pointing out that for $\kappa=-1$ and $|z|=1$, we recover the $\mathcal{N}=2$ $AdS_5$ MN solutions, while for $\kappa=-1$ and $z=0$ we find the $\mathcal{N}=1$ MN solutions. Thus, our results constitute an extension of the MN solutions to an infinite family of $AdS_5$ fixed points preserving at least four supercharges for a fixed $\cC_g$.\footnote{Some of these solutions were also discussed in \cite{Cucu:2003bm,Cucu:2003yk}.}

\subsection{Holographic RG flows} \label{subsec: holoRGflows}

\begin{figure}[t]
\centering
\includegraphics[width=4.75cm]{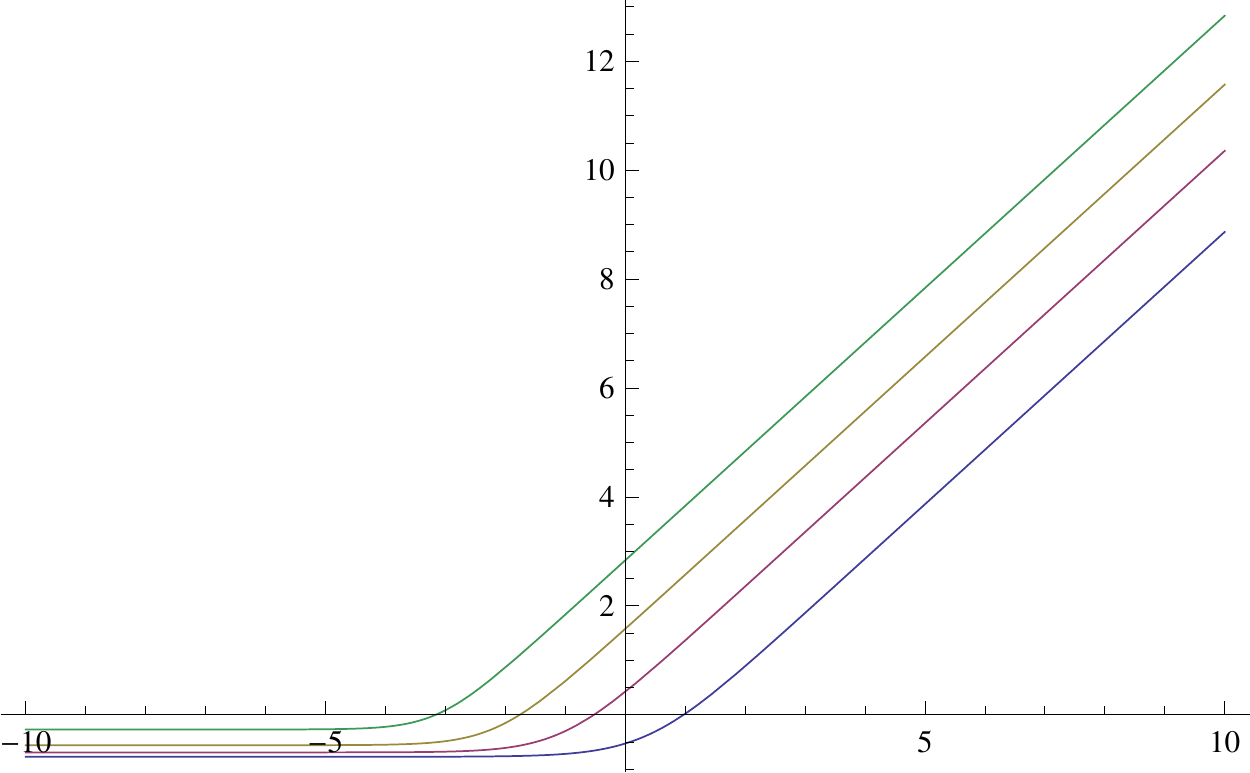}
\hfill
\includegraphics[width=4.75cm]{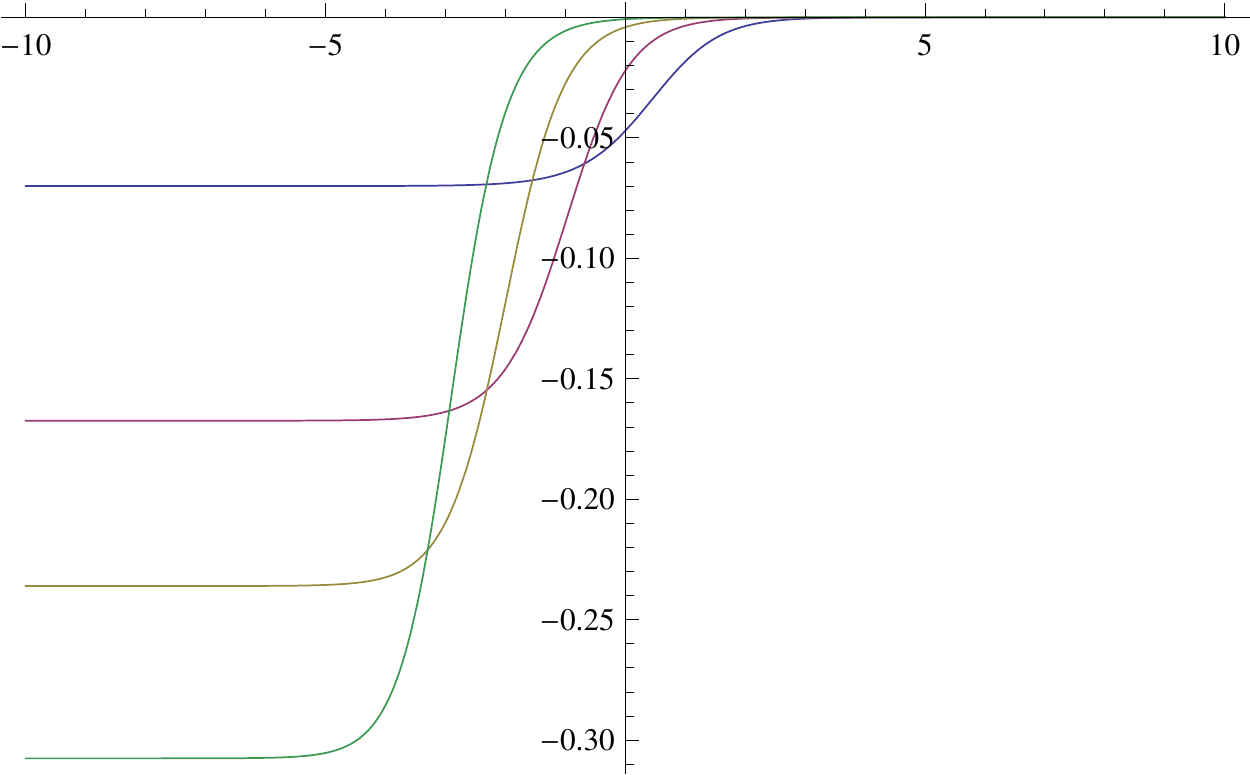}
\hfill
\includegraphics[width=4.75cm]{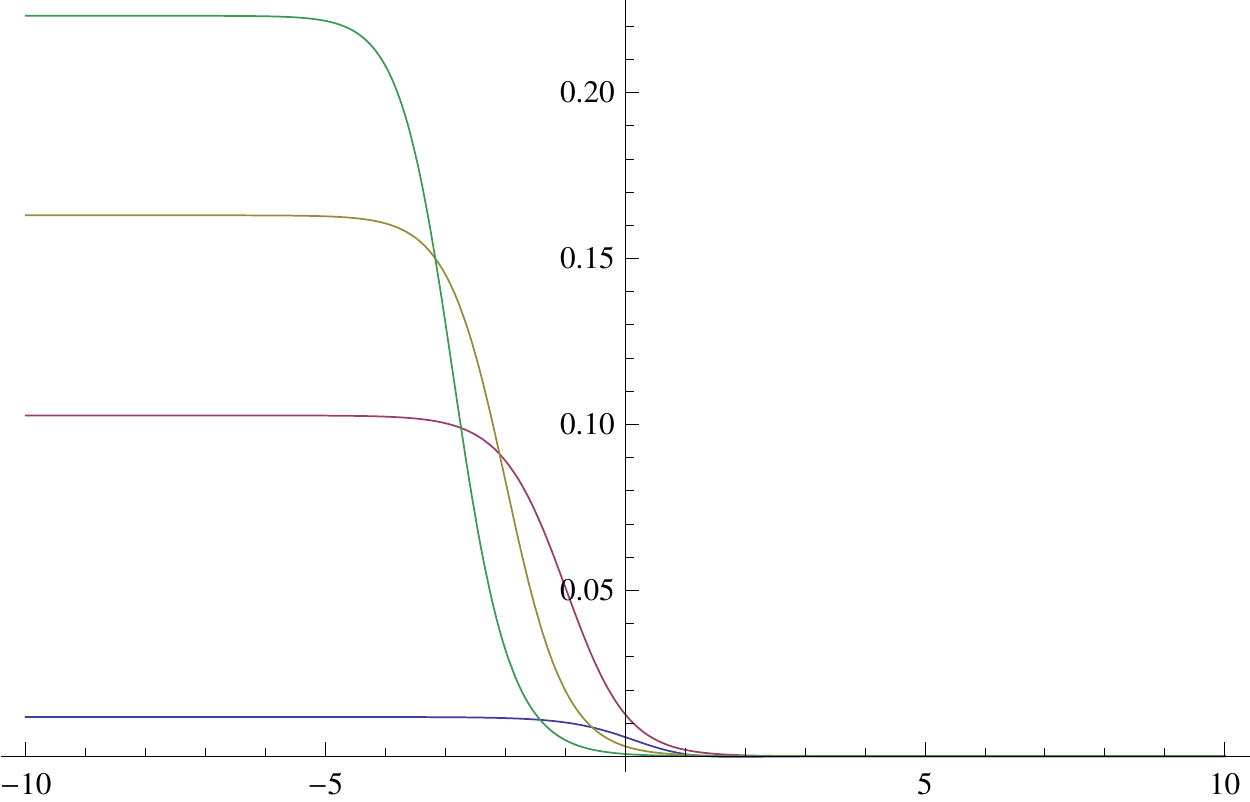}
\caption{{\it Numerical solutions for $g(\rho)$, $\lambda_1(\rho)$ and $\lambda_2(\rho)$ (from left to right). We have fixed $\kappa=-1$ and have chosen four representative solutions for $g=7$ and $z=\{\tfrac{1}{6},\tfrac{2}{3},\tfrac{4}{3},\tfrac{10}{3}\}$.}}
\label{flowplots}
\end{figure}

In this section, we demonstrate the existence of holographic RG flows which interpolate between an asymptotically locally $AdS_7$ region, as described below \eqr{M5gaugescalarAnsatz}, and the IR fixed points discussed above. To do this we restrict to the constant curvature metric on $\cC_g$ and solve the BPS equations \eqref{BPSeqns} with appropriate UV and IR boundary conditions. The BPS equations are nonlinear, and so we resort to numerical methods to find the solutions. It will be convenient to define a new radial variable,
\be
\rho = \frac{1}{5}\left(f(r)+\lambda_1(r)+\lambda_2(r)\right)~,
\ee
in terms of which the equations \eqref{BPSeqns} can be rewritten as
\bea\label{BPSsimple}
 g' &=& 1 + 2e^{6\lambda_1+4\lambda_2}+2e^{4\lambda_1+6\lambda_2} + \ds\frac{\kappa}{2} e^{2\lambda_1+2\lambda_2-2g}((1+z)e^{2\lambda_2}+(1-z)e^{2\lambda_1})~, \notag\\
\lambda_1' &=&2 - 6e^{6\lambda_1+4\lambda_2}+4e^{4\lambda_1+6\lambda_2} + \ds\frac{\kappa}{8} e^{2\lambda_1+2\lambda_2-2g}(3(1+z)e^{2\lambda_2}-2(1-z)e^{2\lambda_1})~,\qquad\\
\lambda_2' &=& 2 + 4e^{6\lambda_1+4\lambda_2}-6e^{4\lambda_1+6\lambda_2} - \ds\frac{\kappa}{8} e^{2\lambda_1+2\lambda_2-2g}(2(1+z)e^{2\lambda_2}-3(1-z)e^{2\lambda_1})~, \notag
\eea
where prime denotes a derivative with respect to $\rho$. We are interested in solutions to these equations which approach the fixed point solutions \eqref{N=1fixedpoints} in the IR ($\rho \to -\infty$) and have the following asymptotic behavior in the UV ($\rho \to \infty$),
\be
g \sim \rho~, \qquad\qquad \lambda_i \sim e^{-10\rho}~.
\ee
This asymptotic behavior is the appropriate one for an asymptotically locally $AdS_7$ solution in the UV. In terms of the original radial coordinate, these boundary conditions are
\be
e^f \sim \ds\frac{1}{r}~, \qquad e^{g} \sim \ds\frac{1}{r}~, \qquad \lambda_i \sim r^2~,
\ee
where the scaling of $\lambda_i$ indicates that a source is turned on for a dimension four operator in the UV theory. One can find numerical solutions to the BPS equations with the above asymptotics for any allowed value of $z$ and both choices of $\kappa$. Numerical solutions for $\kappa=-1$, $g=7$, and a number of representative values for $z$ are presented in Figure \ref{flowplots}. Similar solutions exist for $\kappa=1$.

\section{Holographic duals in eleven dimensions}
\label{sec:eleven}

\subsection{The uplifted solutions}
\label{subsec:trivialuplift}

Since the seven-dimensional supergravity theory with which we work is a consistent truncation of eleven-dimensional supergravity, all solutions discussed in the previous section can be uplifted to eleven dimensions. For this purpose, we utilize the explicit uplift formulae derived in \cite{Cvetic:1999xp}. The eleven-dimensional solution is presented most succinctly in terms of functions $X_{\alpha}$,
\be
X_1 \equiv e^{2\lambda_1}~, \qquad\qquad X_2 \equiv e^{2\lambda_2}, \qquad\qquad X_0 \equiv \left(X_1X_2\right)^{-2}~.
\ee
The metric takes the form
\begin{align}
\begin{split}
&ds^2_{11}= {\Delta}^{1/3} ds_7^2 +\ts\frac{1}{4}\Delta^{-2/3}ds^2_{4}~,\\[5pt]
&ds_7^2= e^{2f}\left(\ds\frac{-dt^2+d\vec{z}^{\,2}+dr^2}{r^2} \right)+ e^{2\hat{g}} (dx_1^2+dx_2^2)~,\\
&ds^2_4=X_0^{-1} d\mu_0^2 +\ds\sum_{i=1}^{2}X_{i}^{-1} (d\mu_i^2+\mu_i^2(d\phi_i+4A^{(i)})^2)~,
\label{11dmetric}
\end{split}
\end{align}
where $A^{(i)}$ are the two seven-dimensional gauge fields \eqref{M5gaugescalarAnsatz} and we have defined
\bea
&\Delta \equiv \ds\sum_{\alpha=0}^{2} X_{\alpha} \mu_{\alpha}^2~, \qquad\qquad \ds\sum_{\al=0}^{2}\mu_{\al}^2 = 1~.
\eea
The periods of the angular coordinates $\phi_i$ are $2\pi$, and as is standard in parameterizing the four-sphere, we have
\be
\mu_0 =\cos\alpha~, \qquad \qquad \mu_1 = \sin\alpha\cos\beta~, \qquad\qquad \mu_2 = \sin\alpha\sin\beta~.
\ee
The eleven-dimensional metric is a warped product of $AdS_5$ with a six-dimensional compact manifold which is a squashed $S^4$ fibration over $\cC_g$. The isometry of the internal manifold for generic values of $\lambda_i$ is $U(1)\times U(1)$.

The four-form flux of the eleven-dimensional solution is given by 
\begin{align}
\begin{split}
 *_{11}F_{(4)}= 4\ds\sum_{\alpha=0}^{2} (X_{\alpha}^2 \mu_{\alpha}^2 -\Delta X_{\alpha})\epsilon_{(7)} &+ 2 \Delta X_{0} \epsilon_{(7)} \\
 &+\ts\frac{1}{4} \ds\sum_{i=1}^{2} X_{i}^{-2} d(\mu_i^2) \wedge (d\phi_i+4A^{(i)}) \wedge *_{7} F^{(i)}~,
\end{split}
\end{align}
where $F^{(i)}=dA^{(i)}$, $\epsilon_{(7)}$ is the volume form for $ds^2_7$, and $*_7$ and $*_{11}$ denote the Hodge star operators for $ds^2_{7}$ and $ds^2_{11}$ in \eqr{11dmetric}, respectively.

\subsection{Central charges and M2 brane operators}

\begin{figure}[t]
\begin{center}
\includegraphics[width=8.5cm]{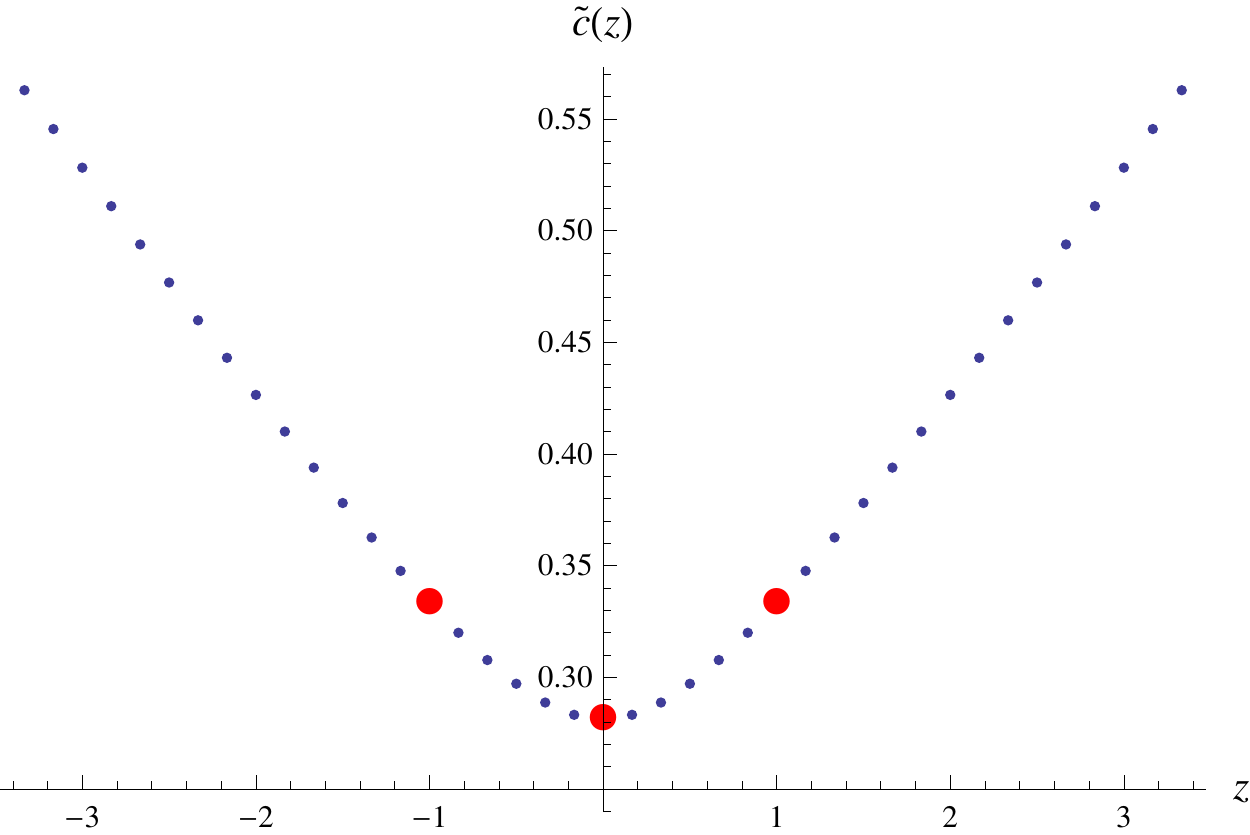}
\caption{\it The central charge $\tilde{c}$ as a function of $z$ for $\kappa=-1$ and $g=7$.}
\label{cplot}
\end{center}
\end{figure}

Given the eleven-dimensional $AdS_5$ supergravity solutions, we can use standard holographic techniques to compute the central charges of the dual field theory \cite{Henningson:1998gx}. As expected for a field theory with a conventional $AdS$ dual, the central charges $a$ and $c$ are equal, and are given by
\be
a=c = \ds\frac{8(1-g)}{3\kappa}e^{2g_0+3f_0}N^3~.
\ee
Using the expressions in \eqref{N=1fixedpoints}, we find
\be
a=c = \ds\frac{(1-g)}{\kappa}N^3\left(\ds\frac{\kappa- \kappa9z^2+(1+3\, z^2)^{3/2}}{48 \,z^2}\right)~, \label{ccuplift}
\ee
matching the result \eqr{aclargeNanomaly} of the anomaly analysis in Section \ref{sec:six}. To parameterize the dependence of the central charge on $z$ it is useful to define
\be
\tilde{c}\equiv \ds\frac{\kappa-\kappa 9z^2+(1+3\, z^2)^{3/2}}{48 \,z^2}~.
\ee
A few comments are in order. The $\mathcal{N}=2$ MN fixed point is at $z=\pm 1$, where one finds $\tilde{c}=1/8$. The $\mathcal{N}=1$ MN solution is at $z=0$, where $\tilde{c}=27/256$, reproducing the famous $27/32$ ratio of these two central charges \cite{Maldacena:2000mw,Tachikawa:2009tt}. It is also instructive to expand the function $\tilde{c}$ for large values of $z$. The result is
\be
\tilde{c}  \sim \ds\frac{\sqrt{3}}{16}\, z-\kappa\ds\frac{3}{16} + \mathcal{O}(z^{-1})~, \qquad |z| \gg 1 ~.
\ee
A plot of the function $\tilde{c} (z)$ is presented in Figure \ref{cplot}. 

Because these supergravity backgrounds are ``wrapped brane geometries'' \cite{Gauntlett:2006ux}, the curve $\cC_g$ is a supersymmetric cycle, and there is a canonical BPS operator which corresponds to an M2 brane wrapping the cycle \cite{Gaiotto:2009gz}. The dimension of this operator is given by the energy of the wrapped M2 brane, which is readily computable in the gravity dual at leading order in $N$. The result is
\be
\Delta(\cO_{M2}) = \ds\frac{8(1-g)}{\kappa} e^{f_0+2g_0-2(\lambda_1^{(0)}+\lambda_2^{(0)})} N = (g-1) N \left(1- \ds\frac{\kappa}{2}\ds\sqrt{1+ 3 z^2}\right)~.
\label{dimTheta}
\ee
For the MN solutions, we reproduce the known results \cite{Gaiotto:2009gz,Benini:2009mz},
\be
\Delta(\cO_{M2})_{|z|=1} = 2 (g-1) N~, \qquad\qquad \Delta(\cO_{M2})_{z=0} = \ds\frac{3}{2} (g-1) N~.
\ee
The expression \eqref{dimTheta} is accurate at large $N$, while we will see in Section \ref{sec:four} that the exact field theory result has $\cO(1)$ corrections.

\subsection{Marginal deformations}

The gravity dual also allows, in principle, for an identification of the exactly marginal deformations of the theory. A definitive treatment of this question would involve finding the linearized supergravity spectrum around each of our solutions, and then trying to find finite deformations of the solution which extend these linearized ones. This is a daunting task which we will not attempt. However, there is a natural set of deformations which are exactly marginal and which we believe to exhaust the list and span the superconformal manifold of the dual field theory. 

The first set of such modes are the infinitesimal variations of the complex structure of $\cC_g$ which enter the construction in the action of the Fuchsian subgroup $\Gamma$ on the hyperbolic plane $\bH^2$ spanned by $(x_1,x_2)$ for $\kappa=-1$ (for $\kappa=1$, there are no such deformations, while for $\kappa$=1 there is one). For $g>1$, there are $3g-3$ such complex deformations, and in the work of \cite{Gaiotto:2009gz} these deformations are identified with exactly marginal gauge couplings of the dual field theory.

The other set of modes correspond to the freedom to shift the gauge fields by a flat connection on $\cC_g$,
\be\label{flatshift}
A^{(1)}\rightarrow A^{(1)}+A_{\rm flat}~,\qquad\qquad A^{(2)}\rightarrow A^{(2)}-A_{\rm flat}~.
\ee
Such a shift leaves the BPS equations satisfied. The space of flat $U(1)$ connections has complex dimension $g$. We thus predict that for generic $g$ and $z$, the number of exactly marginal deformations for these solutions, \ie, the complex dimension of the superconformal manifold of the dual field theories, is $4g-3$. As we show in Section \ref{sec:four} this number is matched by the field theory analysis.\footnote{Note that this applies even for $|z|=1$, corresponding to the $\cN=2$ MN twist. The Abelian Wilson lines are frequently ignored in this case because they preserve only $\cN=1$ supersymmetry. The $3g-3$ complex structure moduli preserve the full $\cN=2$ supersymmetry.}

The non-generic cases are $g=0$, $g=1$, and $z=0$. For $g=0$ there are no complex structure deformations or nontrivial flat connections, and so we predict that these theories are isolated. For $g=1$ there is a single complex structure modulus, and we find $\dim_{\bC}\cM_{g=0}=2$. Finally, for $z=0$ there is an $SU(2)$ structure which was discussed in \cite{Benini:2009mz}, leading to additional marginal deformations related to flat $SU(2)$ connections on $\cC_g$. Consequently, the number of marginal directions for $z=0$ is $6g-6$.

\subsection{Fixed point solutions in canonical coordinates}\label{subsec:canonical}

In \cite{Gauntlett:2004zh}, a set of differential conditions were derived which are obeyed by the metric functions and  fluxes of any quarter-BPS $AdS_5$ solution of eleven-dimensional supergravity. The solutions of the previous section should fall into this general classification, and it is interesting to rewrite them in the corresponding canonical coordinates. One of the benefits of having solutions in these coordinates is that the Killing vector that generates the isometry dual to the superconformal R-symmetry is made manifest. Although there should be a coordinate transformation that brings the solutions of Section \ref{subsec:trivialuplift} into the canonical form, we find it more convenient to directly solve the equations of \cite{Gauntlett:2004zh} with an appropriate Ansatz. 

In this section we present the salient features of the solutions in these coordinates. Details of the derivation are presented in Appendix \ref{a:GMSW}.\footnote{Our focus here is on the case $g\neq 1$. The solutions with $g=1$ are discussed in Appendix \ref{a:GMSW}.} The metric is given by
\be
ds^2_{11} = e^{2\lambda(y,q)} \left[ ds^2_{AdS_5} + e^{2\nu +2A(x)} (dx_1^2+dx_2^2)\right] +  e^{-4\lambda(y,q)}ds^2_{M_4}~, 
\ee
where $ds^2_{AdS_5}$ is the metric of $AdS_5$ with unit radius, and $A(x)$ is the conformal factor for the constant curvature metric on the curve $\cC_g$ with Gaussian curvature $\kappa=\pm1,0$. The squashed four-sphere metric is now written as
\begin{multline}
ds^2_{M_4} = \left( 1+ \ds\frac{4 y^2}{q f(y)} \right) dy^2 + \ds\frac{ f(y) q}{k(q)} \left(dq + \ds\frac{12  y k(q)}{f(y) q} dy\right)^2 \\ 
+ \ds\frac{ a_1^2}{4}~ \ds\frac{f(y)k(q)}{q}(d\chi +V)^2 + \ds\frac{q f(y)}{9} (d\psi + \rho)^2~,
\label{4metricnicebody}
\end{multline}
with $U(1)\times U(1)$ isometry generated by the Killing vectors $\del_\psi$ and $\del_\chi$. The various metric functions are defined as
\begin{align}
\begin{split}
e^{6\lambda(y,q)} &= q f(y) +4y^2~,\\
f(y) &= 1+ 6 (\kappa e^{-2\nu}-6) y^2~,\\
k(q) &= (\kappa e^{-2\nu}-6) q^2 + q - \ds\frac{1}{36}~,
\end{split}
\end{align}
while the one-forms which determine the fibration of the $\psi$ and $\chi$ directions are given by
\begin{align}
\begin{split}
\rho &= (2-2g) V - \ds\frac{1}{2}(a_2 + \ds\frac{a_1}{2q}) (d\chi + V)~,\\
dV &= \ds\frac{\kappa}{2-2g} e^{2A} dx_1 \wedge dx_2~.
\label{oneforms}
\end{split}
\end{align}
The various constants take values
\be
a_1 \equiv \ds\frac{2(2-2g)e^{2\nu}}{\kappa }~, ~~~ a_2 \equiv - 6a_1 \left(1 - \ds\frac{\kappa}{6e^{2\nu}}\right)~,~~~e^{2\nu} = \ds\frac{1}{6} ( -\kappa \pm \sqrt{1 +3 z^2} )~.
\ee
The coordinates $q$ and $y$ have finite range determined by the zeroes of the functions $k(q)$ and $f(y)$,
\begin{align}
\begin{split}
&q \in \left[q_-,~q_+\right]~, \qquad\qquad q_{\pm}^{-1} = 18 \pm 6 \ds\sqrt{3 + \ds\frac{2\kappa + 2\sqrt{1 + 3 z^2}}{\kappa z^2}}~, \\
&y\in \left[y_{-},~ y_{+}\right]~, \qquad\qquad y_{\pm}^{-1} = \pm 6 \sqrt{1+ \ds\frac{\kappa}{\kappa-\sqrt{1+3z^2}} }~.
\end{split}
\end{align}
The four-form flux can also be written down explicitly and is presented in Appendix \ref{a:GMSW}. 

There are a number of notable features of these backgrounds. The solutions are parametrized by the twist parameter $z$, and the internal manifold is an $S^4$ fibration over a closed Riemann surface, of any genus, with $U(1)\times U(1)$ isometry. With the exception of the $\kappa=-1$, $z=0$ solution, this six-manifold is {\it not complex}, in contrast to the solutions found in \cite{Gauntlett:2004zh}. The Killing vectors are expressed in terms of the more natural Killing vectors $\del_{\phi_1}$ and $\del_{\phi_2}$ of \eqr{11dmetric} (which have compact integral curves) as
\be
\del_{\chi}=\del_{\phi_1}-\del_{\phi_2}~,\qquad\qquad
\del_{\psi}=(\del_{\phi_1}+\del_{\phi_2})+\epsilon(\del_{\phi_1}-\del_{\phi_2})~.
\label{anglechange}
\ee
Here $\epsilon$ is as in \eqref{epslargeN}, and so we see that the canonical coordinates automatically know the superconformal R-symmetry of the fixed point theory. It is an open question how to realize the analogue of $a$-maximization (\`a la \cite{Martelli:2005tp,Martelli:2006yb,Eager:2010yu,Gabella:2010cy}) from the supergravity perspective.

As a brief aside, we point out that orbifolds of the supergravity solutions described above provide the holographic duals to several related theories. In particular, by dividing the solution \eqr{11dmetric} by the action
\be
\phi_1\to-\phi_1~,\qquad\phi_2\to-\phi_2~,\qquad\mu_0\to-\mu_0~,
\ee
the topological four-sphere becomes a (smooth) topological $\bR\bP^4$ which is fibered over $\cC_g$. The resulting supergravity solutions are dual to the IR SCFT arising from the twisted compactification of the $D_N$ $(2,0)$ theory on $\cC_g$. The central charges of the dual field theories in the large $N$ limit can be computed and the result is four times larger than the expression in \eqref{ccuplift}. This matches the large $N$ limit of the central charges computed via the anomaly polynomial \eqref{exactcentral} for $G=D_N$.

There is another class of orbifolds which preserves all the supersymmetries of our solutions.\footnote{For $|z|=1$ this orbifold reduces the number of preserved supercharges from eight to four.} The orbifold action is
\be
\phi_1\to\eta\phi_1~,\qquad\phi_2\to\eta^{-1}\phi_2~;\qquad\eta^k=-1~,
\ee
where $k\in \bZ$, which is the action of $\bZ_k\subset U(1)_{\cF}$. This action has fixed points at the North and South poles of the four-sphere, leading to $A_k$-type ADE singularities in the supergravity solution. These solutions are dual to the infrared limit of the theory on $N$ M5-branes wrapped on $\cC_g$ and placed at an $A_k$ singularity. In other words, to get to this IR SCFT we first perform a $\bZ_k$ orbifold on the $A_N$ $(2,0)$ theory to get a $(1,0)$-supersymmetric theory, and then compactify on $\cC_g$ with a partial twist with parameter $z$.

\section{Intermediate theories from generalized quivers}
\label{sec:four}

In this section, we describe four-dimensional field theory constructions of the fixed points with $p,q \geq 0$. These are precisely those fixed points with central charges taking intermediate values between those of the $\cN=1$ and $\cN=2$ MN theories. The primary building block of these dual theories is a strongly coupled, isolated $\cN=2$ SCFT denoted by $T_N$ \cite{Gaiotto:2009we}, which is then coupled to $\cN=1$ and $\cN=2$ vector multiplets. Although the $T_N$ theory has no known Lagrangian description, it is still possible to calculate operator dimensions and central charges via information about global symmetries. In addition to providing useful tools for analyzing the new fixed points, the theories in this section substantially increase the number of known $\cN=1$ SCFTs constructed from the $T_N$ theory, extending the work of \cite{Benini:2009mz,Bah:2011je}.

\subsection{$T_N$ basics}

The $T_N$ theories were first discovered by Gaiotto in \cite{Gaiotto:2009we}, and are the low-energy theories coming from $N$ M5-branes wrapping a thrice-punctured sphere. They are $\cN=2$ SCFTs with $SU(2)_R \times U(1)_R \times SU(N)^3$ global symmetry and no known weakly coupled Lagrangian description (except in the special case $N=2$, when the theory is free). These theories have Higgs branch operators $\mu_a$ with $a = 1,2,3$; each such operator has scaling dimension two and transforms in the adjoint of one $SU(N)$. There are also Coulomb branch operators $u_k^{(i)}$ with $k = 3,...,N$ and  $i=1,...,(k-2)$, and with dimension $ \Delta[u_k^{(i)}]=k$.
Finally, there are dimension $(N-1)$ operators $Q$ and $\widetilde {Q}$ which transform in the $({\bf N},{\bf N},{\bf N})$ and $(\overline {\bf N}, \overline {\bf  N}, \overline {\bf N})$ representations of  $SU(N)^3$. 

Since we are constructing theories with only $\cN=1$ supersymmetry, it is useful to think about $T_N$ from an $\cN=1$ point of view. We will denote by $I_3$ the generator of the Cartan of $SU(2)_R$, and by $R_{\mathcal{N}=2}$ the generator of $U(1)_R$. When expressing this theory as an $\cN=1$ SCFT, the superconformal R-symmetry is generated by
\be
R_{\cN=1} = \tfrac 13 R_{\mathcal{N}=2} + \tfrac 43 I_3~. \label{R1}
\ee
There is also a global symmetry $J$ which commutes with the $\cN=1$ supercharges,
\be
 J = R_{\mathcal{N}=2} - 2 I_3~. \label{J}
\ee
When supersymmetry is explicitly broken to $\mathcal{N}=1$ by coupling to additional matter, $R_{\cN=1}$ will no longer be in the same multiplet as the stress tensor, and the superconformal R-symmetry at an IR fixed point can potentially be a different linear combination of the available $U(1)$ symmetries. For more details on the action of these symmetries and their anomalies see Appendix \ref{a:scftrev}.

\begin{figure}[h]
 \centering
\includegraphics[scale=1.5]{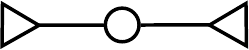}\caption{\it The simplest generalized quiver diagram, showing two $T_N$ theories (represented by triangles, \ie, trinions) connected by a vector multiplet (represented by a circle).}\label{tnvec}
\end{figure}

The $T_N$ theory can be used as an ingredient in generalized quiver gauge theories which are constructed by gauging the $SU(N)$ global symmetries. In particular, by gauging $SU(N)_{\rm diag} \subset SU(N) \times SU(N)$, we can connect different $T_N$ blocks as in Figure \ref{tnvec}. The associated vector multiplet can be either $\cN=1$ or $\cN=2$. In the latter case, we think about this vector multiplet as an $\cN=1$ vector and an $\cN=1$ adjoint chiral multiplet $\phi$. The chiral multiplet comes with a global $U(1)$ symmetry under which it has charge one. Although by itself this $U(1)$ is anomalous, it can combine with other anomalous global symmetries to form anomaly-free symmetries.
We now consider a simple example in which two different $T_N$'s are connected together by gauging a diagonal $SU(N)$ (see Figure \ref{tnvec}). We label the two $T_N$'s by $i = 1,2$, and the three $SU(N)$ flavor symmetries for the $i^{\rm th}$ $T_N$ by $i_a, a=1,2,3$. 

Gauging $SU(N)_{\rm diag} \subset SU(N)_{1_a} \times SU(N)_{2_b}$ with an $\cN=1$ vector multiplet leads to a theory which in the UV is a decoupled pair of $T_N$ blocks along with a free vector, and which flows to strong coupling in the infrared. The anomaly-free R-symmetry and the anomaly-free Abelian flavor symmetry are given by
\be\begin{split}
R_0 &= R_{\cN=1} + \tfrac{1}{6} \left(J_1 + J_2 \right)~, \\
\mathcal{F} &= \tfrac 12 (J_1 -J_2)~. \label{n0}
\end{split}
\ee  
Note that $\Tr \cF = 0$, so this global symmetry is ``baryonic". As discussed in \cite{Intriligator:2003jj}, this means that it cannot mix with any candidate superconformal R-symmetry. Thus, when $T_N$'s are connected only with $\cN=1$ vectors, the superconformal R-symmetry at an IR fixed point is simply $R_0$.

Now, consider instead connecting the $T_N$'s with an $\cN=2$ vector multiplet. We label the global $U(1)$ that comes with the $\cN=1$ adjoint chiral multiplet by $F_{1_a 2_b}$. The anomaly-free Abelian symmetries are then
\be
\begin{split}\label{n1}
R_0 &= R_{\cN=1} + \tfrac{1}{6} \left(J_1 + J_2 \right)~, \\
\mathcal{F}_1 &= \tfrac 12 (J_1 + F_{1_a 2_b})~, \\
\mathcal{F}_2 &= \tfrac 12 (J_2 +F_{1_a 2_b})~.
\end{split}
\ee 
The flavor $U(1)$'s are most conveniently described in a different basis,
\be
\begin{split}
\mathcal{F}_+ &= \mathcal{F}_1 + \mathcal{F}_2 = \tfrac 12 (J_1 + J_2 + 2 F_{1_a 2_b})~,\\
\mathcal{F}_- &= \mathcal{F}_1-\mathcal{F}_2 = \tfrac 12 (J_1 -J_2)~.\label{n2}
\end{split}
\ee  
$\cF_-$ is baryonic and will not mix with the R-symmetry. However, $\cF_+$ is not baryonic, and so the superconformal R-symmetry can be of the form $R_0 + \epsilon \cF_+$ for some real $\epsilon$. The value of $\epsilon$ is determined by $a$-maximization, as we will soon discuss.

The $\cN=2$ gauging also introduces a superpotential of the form
\be\label{Wexample}
W = \Tr \phi_{1_a 2_b} (\mu_{1_a} + \mu_{2_b})~.
\ee
From the charge assignments in Appendix \ref{a:scftrev}, we see that this in fact breaks $\cF_-$ explicitly. We will see this phenomenon repeated later in this section, where precisely all baryonic symmetries will generally be broken by superpotential couplings of the form \eqr{Wexample}. This is crucial for matching with the gravity constructions, which generally have $U(1)^2$ isometry and no baryonic symmetries. 

An important outcome of this discussion is that the relative sign appearing in front of $J_i$ in the $U(1)$ flavor symmetry is directly correlated with whether we use $\cN=1$ or $\cN=2$ vector multiplets. When the multiplet is $\cN=2$, the $J_i$ have the same sign, while for an $\cN=1$ multiplet, they have opposite sign.

\subsection{The general quivers}

\begin{figure}[ht]
 \centering
\includegraphics[scale=.8]{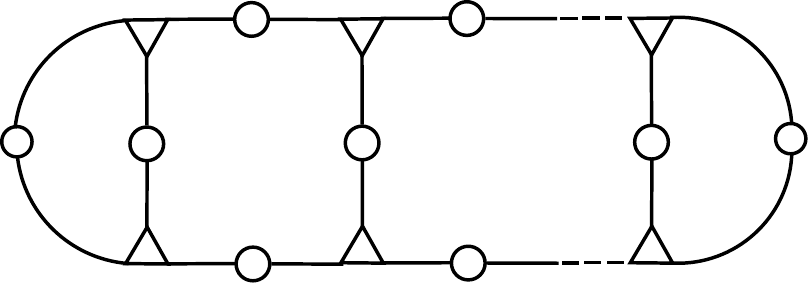}\caption{\it $T_N$'s connected by vector multiplets.}\label{Sicilian}
\end{figure}

The theories we wish to study are constructed as described above, where we now connect $2g-2$ copies of the $T_N$ theory together in such a way that all the $SU(N)$ global symmetries have been gauged, as in Figure \ref{Sicilian}. For each gauging, we can choose either an $\cN=1$ or $\cN=2$ vector multiplet. We want to make these choices so that there is always a single non-baryonic $U(1)$ flavor symmetry preserved, generalizing $\cF_+$ in the above example. 

To do this, we assign a sign $\sigma_i = \pm 1$ to the $i^{\rm th}$ $T_N$, so that each quiver comes with a set of signs $\{ \sigma_i \}$. If two $T_N$ blocks $i$ and $j$ are connected and $\sigma_i \sigma_j = 1$, the gauging uses an $\cN=2$ vector multiplet, while for $\sigma_i \sigma_j = -1$ it uses an $\cN=1$ vector multiplet.\footnote{It is possible to pick assignments of $\cN=1$ and $\cN=2$ vector multiplets in a way which is incompatible with this construction. We believe that such theories will always flow to the $\cN=1$ MN theory.} 

\begin{figure}[h!]
 \centering
\includegraphics[scale=1]{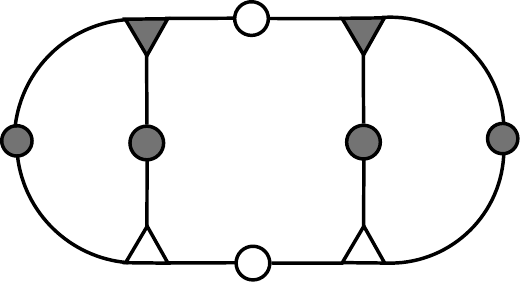}\caption{\it An example of a shaded quiver with $p=q=2$.}\label{shading}
\end{figure}

We will define $p$ to be the number of trinions with $\sigma_i = 1$ and $q$ the number of trinions with $\sigma_i = -1$,  with $p + q = 2g-2$. We also define $m_1$ to be the number of $\cN=1$ gauge groups and $m_2$ the number of $\cN=2$ gauge groups, with $m_1+m_2=3g-3$. Graphically, we can represent the sign of each $T_N$ by shading it in or not. We adopt the convention that for $\sigma = +1$ the $T_N$ gets shaded, while for $\sigma = -1$ the $T_N$ is left unshaded. Similarly, we shade $\cN=1$ vector multiplets but do not shade $\cN=2$ vector multiplets. Given the rules described in the previous paragraph, trinions of the same shading will be connected by unshaded nodes, while trinions with opposite shading will be connected with shaded nodes; see Figure \ref{shading}.

In terms of the Calabi-Yau geometries discussed in Section \ref{sec:six}, this prescription has a simple heuristic interpretation. Each $T_N$ building block corresponds to a three-punctured sphere embedded in a local geometry of the form
\be\label{triniongeometry}
T^*\cC_{0,3}\times\bC~,
\ee
where $\cC_{0,3}$ is the thrice-punctured sphere. When such geometries are glued together to form a Calabi-Yau threefold of the form \eqr{locallines}, the curved part of the normal bundle can be chosen to lie in either of the two line bundles. If it is chosen to be a part of the bundle $\cL_{k_i}$, the sign associated to the $T_N$ is $\sigma_i = (-1)^{k_i+1}$. When two $T_N$'s have the same sign, the normal geometry remains a cotangent bundle under gluing, so the gauging is that of an $\cN=2$ theory. But when the signs are different, the geometry only retains half of its supersymmetry under gluing, leading to an $\cN=1$ gauging in the field theory. In such a gluing construction of the Calabi-Yau threefold geometry, the number of punctured spheres over which a given one of the line bundles looks like the cotangent bundle is equal to the degree of that line bundle. Thus the numbers $p$ and $q$ here are precisely those defined in \eqr{chernnumbers}.

Because the $\cN=1$ gauge couplings have positive beta functions, the theories defined by these generalized quivers are described in the UV by a collection of decoupled $\cN=2$ SCFTs along with free $\cN=1$ vector multiplets. In the infrared, the gauge groups become strongly coupled, and the theories flow to $\cN=1$ SCFTs which are precisely of the type discussed in the previous sections. We now support this claim by analyzing the anomalies and symmetries of these theories. 

One possible anomaly-free R-symmetry for any generalized quiver of this type is given by
\be\label{genR0}
R_0=R_{\cN=1}+\tfrac 16\sum_{i}J_i~.
\ee
This is the analogue of the symmetry $U(1)_K$ which was always present in the geometric constructions of Section \ref{sec:six}. In general $R_0$ will not be the superconformal R-symmetry of an IR fixed point, since it may mix with the other Abelian global symmetry present in our models. Given a choice of signs $\{ \sigma_i \}$, this additional symmetry is given by
\be
\label{fis}
\cF = \tfrac 12 \sum_i \sigma_i J_i +  \tfrac 12 \sum_{\langle ij\rangle} (\sigma_i+\sigma_j) F_{i_a j_b}~.
\ee
In this expression, the last sum is over the locations of all $\cN=2$ vector multiplets. The action of this symmetry on the individual $T_N$ blocks is just right to be identified with the geometrically realized symmetry of the same name in Section \ref{sec:six}. Importantly, $\cF$ is unbroken by superpotential terms of the form $\mu \phi$ at $\cN=2$ nodes and $\mu_i \mu_j$ at $\cN=1$ nodes.

In the absence of accidental symmetries, the superconformal R-symmetry of an IR fixed point will then be of the form
\be\label{RsymFT}
R = R_0 + \epsilon \cF~,
\ee
where $\epsilon$ is a real number which is determined by $a$-maximization \cite{Intriligator:2003jj}. Recall that this procedure singles out the unique R-current in the same multiplet as the stress tensor by maximizing the trial central charge as a function of $\epsilon$,
\be
\label{aofs}
 a(\epsilon) = \tfrac 3{32} \left ( 3 \mbox{Tr} R^3(\epsilon) - \mbox{Tr} R(\epsilon) \right ).
\ee  

It is straightforward to compute the contribution of both the $T_N$ blocks and the vector multiplets to the trial central charge. In particular, the contribution of the $i$'th $T_N$ is
\be \begin{split}
 a_{T_N}(\epsilon) &= \tfrac{3}{32} A(\sigma_i \epsilon,N)~, \\
A(\epsilon,N) & \equiv \left[\tfrac 38 \left(1 + \epsilon\right)^3 - \tfrac 12 \left(1 + \epsilon\right)\right] \mbox{Tr}R_{\mathcal{N}=2}^3  + \tfrac 92 \left(1 + \epsilon\right)\left(1-\epsilon\right)^2 \mbox{Tr} R_{\mathcal{N}=2}I_3^2~. 
\end{split}
\ee
The contribution of an $\cN=1$ vector multiplet is 
\be
 a_{i_a j_b} =\tfrac{6}{32}(N^2 -1) ~,
\ee 
while that of an adjoint chiral multiplet is
\be
 a_{i_a j_b}(\epsilon) =\tfrac{3}{32}(N^2 -1) \left[\tfrac 38 (\sigma_i + \sigma_j)^3 \epsilon^3 -\tfrac 12  (\sigma_i + \sigma_j)\epsilon \right]~.
\ee 
Summing up the contributions from the entire quiver, we then need to maximize
\be
 a(\epsilon) = \tfrac{3}{32} \left[p A(\epsilon,N) + q A(-\epsilon,N) \right] + \tfrac{3}{32} (N^2-1)\left[3 (p+q) + \tfrac 32(p-q) (3 \epsilon^3 -\epsilon) \right]~.\label{cctrial}
\ee
The results exactly reproduce the values of $\epsilon$, $a$, and $c$ from Equations \eqr{maxepsilon} and \eqr{exactcentral} for the choices $G=A_{N-1}$ and $\kappa=-1$.

\subsection{Operator dimensions and the conformal manifold}

Given the R-charges, we can find the exact scaling dimensions of the various chiral primary operators using $\Delta = \frac32 R$. The operators $\mu$, $\phi$, $Q$, and $u_k$ have dimensions
\be\label{opdims}
\begin{split}
\Delta[\mu] &= \tfrac 32 \left (1- \sigma_i {\epsilon} \right )~, \qquad\qquad\qquad\quad\;\, \Delta[\phi_{i j}]= \tfrac 32(1 + \tfrac 12 (\sigma_i + \sigma_j) {\epsilon}) ~, \\
\Delta [Q] &= \tfrac34 (N-1) \left (1-   \sigma_i {\epsilon} \right)~,\qquad\qquad \Delta[u_k]=\tfrac 32\left(1+\sigma_i \epsilon\right)k~.
\end{split}
\ee
One might worry that some of these operators could violate the unitarity bound, but since $0 \leq |\epsilon| \leq 1/3$ for $p,q\geq 0$, they do not. 

All relevant and marginal operators can be constructed solely out of $\mu$ and $\phi$. For the purpose of enumerating these operators, from now on we assume without loss of generality that $p\leq q$. At an $\cN=2$ node, the operators $\Tr \mu_{i_a}\phi_{i_a j_b}$ and $\Tr \mu_{j_b}\phi_{i_a j_b}$ are marginal. If that node connects shaded $T_N$'s, then the operators $\Tr \mu_{i_a}^2$ and $\Tr{\mu_{j_b}^2}$ are relevant. Alternatively, if the node connects unshaded $T_N$'s, then the only relevant operator is $\Tr \phi_{i_a j_b}^2$. At an $\cN=1$ node, there is a single marginal operator $\Tr \mu_{i_a}\mu_{j_b}$ and a single relevant operator $\Tr \mu_{i_a}^2$, where $i$ labels the shaded $T_N$. 

Additionally, we can construct gauge invariant operators out of the tri-fundamentals $Q$ and $\tilde{Q}$. These operators correspond to the wrapped M2-brane operator of Section \ref{sec:eleven},
\be
 \mathcal{O}_{M2} = \prod_{i=1}^{2g-2} Q_i~, \qquad\qquad \widetilde{\mathcal{O}}_{M2} = \prod_{i=1}^{2g-2} \widetilde{Q}_i~,
\ee  
with dimensions
\be
\Delta[\mathcal{O}_{M2}]= \Delta[\widetilde{\mathcal{O}}_{M2}] =\frac 34 (N-1) \left( (p+q) -{\epsilon} (p-q) \right)~.
\ee
For large $N$, these dimensions match the result from M2-branes wrapping $\cC_g$, as described by equation \eqr{dimTheta}.

We can compute the dimension of the conformal manifold via the method of Leigh and Strassler \cite{Leigh:1995ep} (or equivalently \cite{Green:2010da}). For a quiver construction with $m_1$ $\cN=1$ vector multiplets and $m_2$ $\cN=2$ vector multiplets, there are $m_1+2 m_2$ marginal operators. Additionally, there are $3g-3$ marginal gauge couplings, giving a total of $3g - 3 + m_1+2m_2$ marginal deformations. However, there are constraints on the anomalous dimensions coming from each of the $2g-2$ $T_N$'s and the $m_2$ $\phi_{i_a j_b}$, with the exception of one overall linear combination. Thus, the total number of constraints is $2g-3 + m_2$, and the complex dimension of the conformal manifold is given by
\be
\dim_{\bC}\cM_{\rm C} = m_1 + m_2 + g = 4g-3~.
\ee
This matches the counting of Section \ref{sec:eleven} exactly.

\section{An example: genus three}\label{sec:example}

\begin{figure}[ht!]
\centering
\includegraphics[scale=.65]{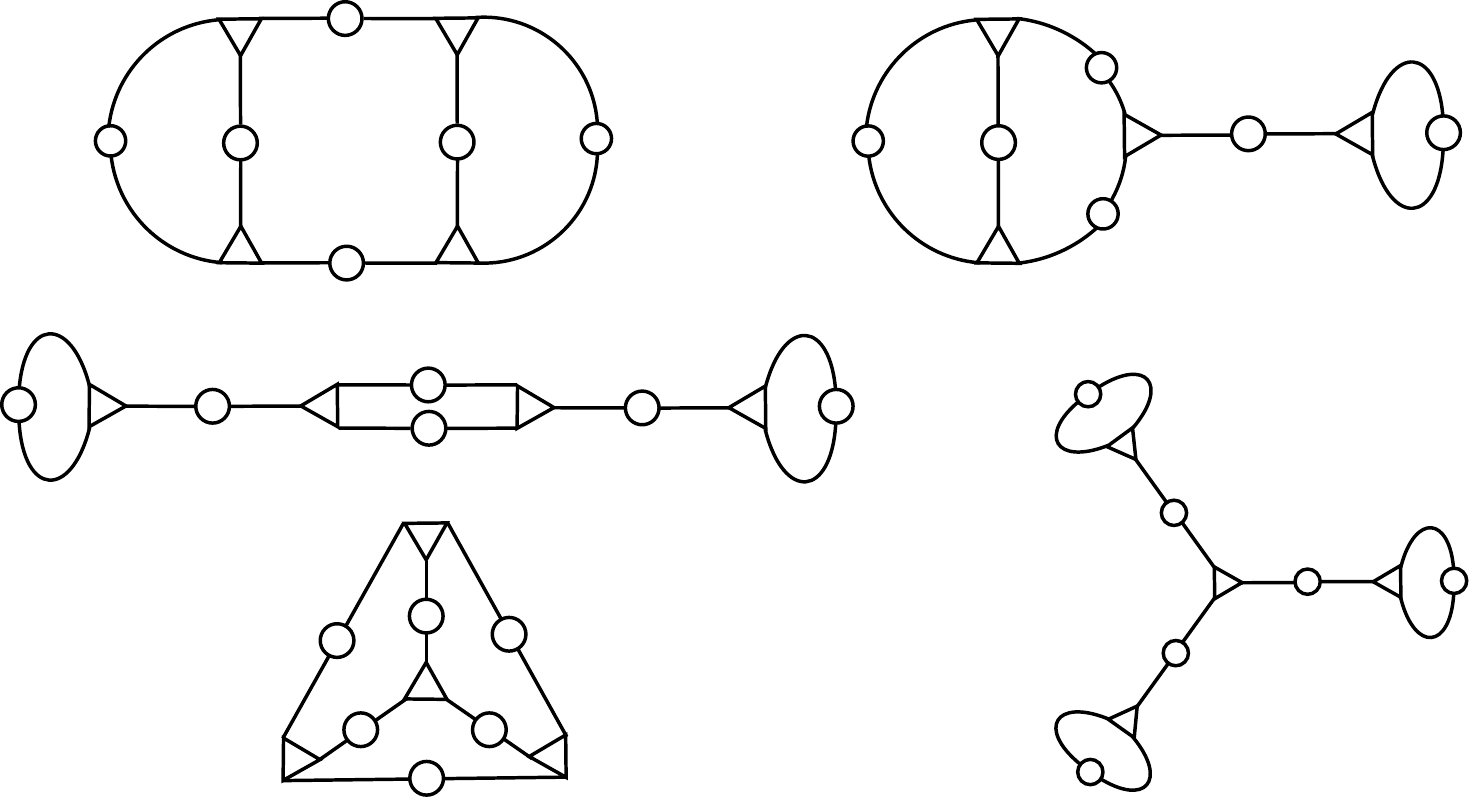}\caption{\it The five different quiver topologies for a genus 3 Riemann surface.}\label{allfour}
\end{figure}

To illustrate the above procedure for building these theories, we now work through the details of the genus three case. The quivers for this theory have $2g-2 = 4$ $T_N$'s and $3g-3 = 6$ gauge groups. It is sufficient to consider $p=0,1,2$ since  the theories with $p=3,4$ are equivalent to the ones with $p=1,0$, respectively. There are five different quiver topologies, which are displayed in Figure \ref{allfour}. 

For any given quiver topology, we can choose any one of $p=0,1,2$. All $g=3$ theories with the same value of $p$ should flow to the same fixed point in the infrared, although as we will see,  there are some puzzles associated with this claim. In particular, we will see that the spectrum of gauge-invariant chiral operators does not na\"ively match between the different UV starting points. How to properly account for this matching remains unclear, and probably relies on chiral ring relations for $T_N$ theories that we do not yet fully understand. We leave the resolution of this discrepancy for future work.

\begin{itemize}

\item $p=0$

In this case, all gauge fields belong to $\cN=2$ vector multiplets and the theory is dual to the $\cN=2$ MN solution. The R-symmetry is as in \eqref{RsymFT} with $\epsilon=1/3$. This theory has been analyzed in great detail in \cite{Gaiotto:2009gz}, so we keep our discussion brief. One important difference between this case and the purely $\cN=1$ constructions is that all the nonzero (gauge) couplings are exactly marginal. All quiver topologies are S-dual to one another, as argued in \cite{Gaiotto:2009we}. 

The $\cN=1$ perspective also introduces some new features into the analysis of these theories. In particular, the discussion of the previous section goes through and the number of marginal deformations which preserve at least $\cN=1$ supersymmetry is $4g-3=9$. Of these, $3g-3$ are identified with the exactly marginal gauge couplings which preserve the full $\cN=2$ supersymmetry. The additional $g=3$ deformations are superpotential deformations which break $\cN=2\to\cN=1$. These superpotential deformations are modifications of \eqr{Wexample} so that the two terms coming from a given node have different coefficients. Finally, it is clear from \eqref{opdims} that there are six relevant operators for any realization of the quiver which are mass terms for the chiral superfields. Deformation by these mass terms leads to the $\cN=1$ MN theories \cite{Benini:2009mz}.

\item $p=1$

Up to equivalence, this is the only intermediate theory for $g=3$. For the moment, we focus on the UV quiver illustrated in Figure \ref{g3p1}, in which none of the $T_N$'s connect to themselves. Since $p=1$, we have picked one of the $T_N$'s to shade. Because the quiver has $D_4$ symmetry, all such choices are equivalent, so without loss of generality we choose the one in the lower left corner. The remaining $T_N$'s are not shaded, so all nodes connected to the shaded $T_N$ are $\cN=1$ vector multiplets. The other three nodes in this quiver are $\cN=2$ vector multiplets. 

\begin{figure}[t]
\centering
\includegraphics[scale=1.2]{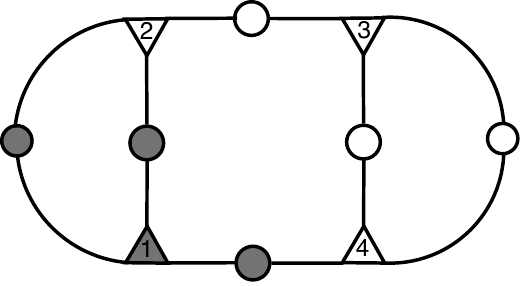}\caption{\it A quiver with $g=3$ and $p=1$.}\label{g3p1}
\end{figure}

The global flavor symmetry $\cF$ defined in \eqref{fis} is given by
\be
\cF = \tfrac 12 \left(J_1 - J_2 - J_3 - J_4\right) - F_{2_1 3_1} - F_{3_24_2} - F_{3_34_3}~,
\ee
where the sign in front of each $J_i$ is correlated with the shading, and the $F_{i_ai_b}$ symmetries are only present at $\cN=2$ nodes. The R-symmetry is determined by \eqref{RsymFT} with
\be
\epsilon = \ds\frac{2N(1-N^2)+\sqrt{7N^6-10N^4-4N^3+4N^2+2N+1}}{3(N^3-1)}~.
\ee

The remaining quivers with $p=1$ provide different UV realizations of the same IR SCFT and are given in Figure \ref{allp1}. It is informative to compare the UV quiver in Figure \ref{g3p1} with another realization of the same theory; for illustrative purposes we pick the theory in the upper left corner of Figure \ref{allp1}. As discussed at the end of Section \ref{sec:four}, the complex dimension of the conformal manifold for both of these theories is $\dim_{\bC}\cM_{\rm C} = 4g-3 = 9$. However we will now show that they na\"ively appear to have a different number of relevant operators.

\begin{figure}[ht]
\centering
\includegraphics[scale=.55]{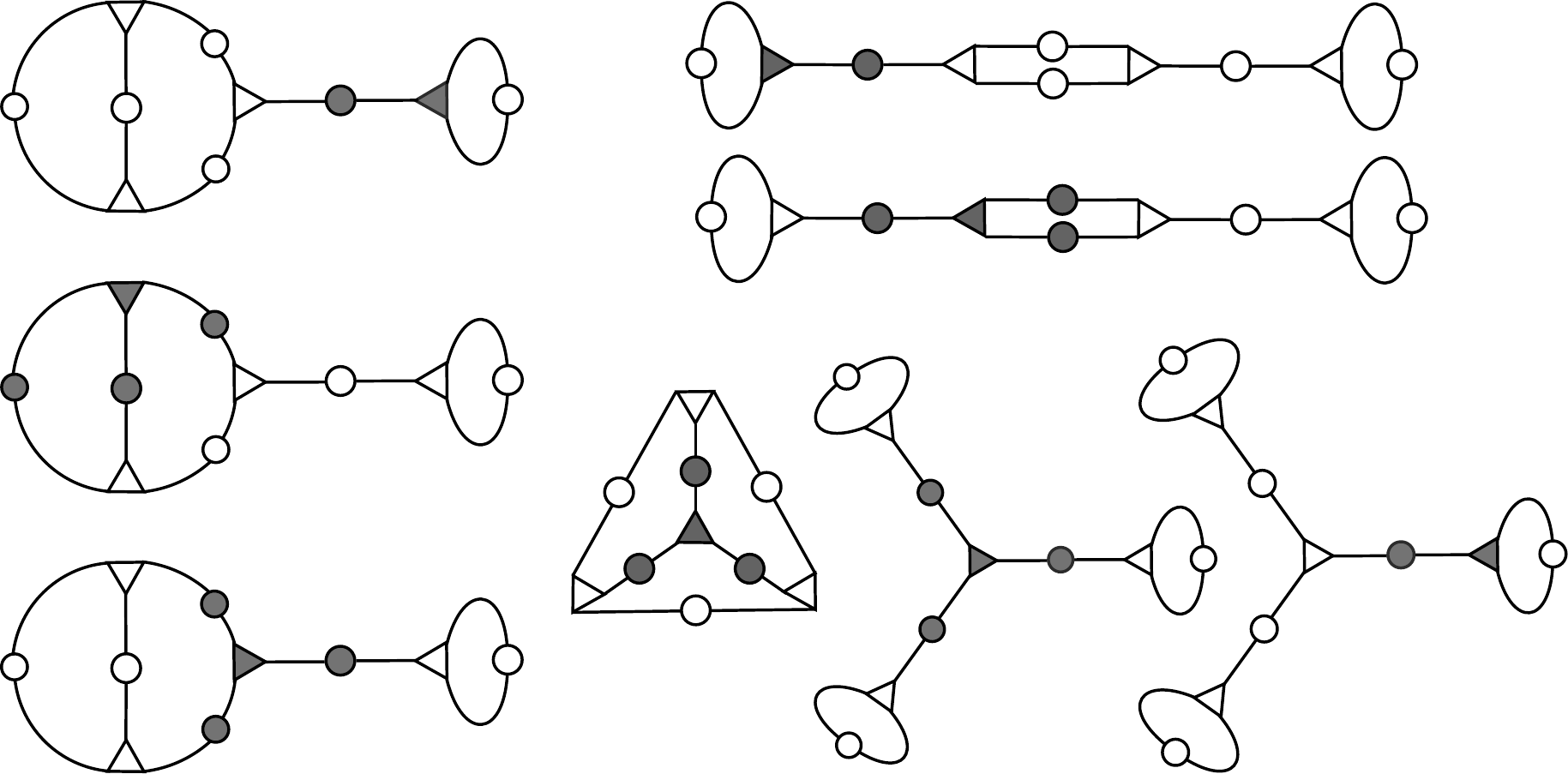}\caption{\it The remaining quivers with $g=3$ and $p=1$.}\label{allp1}
\end{figure}

Using the expressions for operator dimensions in \eqref{opdims} we count four relevant operators of dimension $\Delta=3(1-\epsilon)$ in the theory of Figure \ref{g3p1},
\be
\Tr \phi_{2_13_1}^2~, \qquad \Tr \phi_{3_24_2}^2~, \qquad  \Tr \phi_{3_34_4}^2~, \qquad \Tr \mu_{1}^2~,
\ee
where $\Tr \mu_1^2$ represents the three operators $\Tr \mu_{1_a}^2$ which are believed to be equivalent in the chiral ring of the $T_N$ theory \cite{Benini:2009mz}. For the quiver in the upper left of Figure \ref{allp1}, we label the unshaded $T_N$'s with $1,2,3$ counterclockwise and the shaded one by $4$. There are then five relevant operators (with the same dimension as before) given by
\be
\Tr \phi_{1_12_1}^2~, \quad \Tr \phi_{1_22_2}^2~, \quad  \Tr \phi_{1_33_1}^2~, \quad  \Tr \phi_{2_33_3}^2~, \quad \Tr \mu_{4}^2~.
\ee
As presented, the spectrum of gauge-invariant operators appears not to match between the two theories. Nevertheless, these two theories have the same central charges and their conformal manifolds are of equal dimensions. Moreover, we have given evidence in Section \ref{sec:six} that they arise from the same M5-brane construction, so we are led to conjecture that this discrepancy should be resolved. A natural guess is that there are nontrivial chiral ring relations for the $T_N$ theories which we have not taken into account which will modify the counting of chiral operators. 

\item $p=2$

This theory is dual to the $\cN=1$ MN solution. There are two different realizations of the quiver with topology as in Figure \ref{g3p1}, since there are two inequivalent ways to shade the trinions. These are illustrated in Figure \ref{bothp2}. The quiver on the left has only $\cN=1$ vector multiplets, and is the same as the quiver studied in \cite{Benini:2009mz}. Because $\epsilon = 0$, the R-symmetry of this theory is simply $R_0$ as defined in \eqref{genR0}. This is the special theory for which there are additional marginal deformations, and it was argued in \cite{Benini:2009mz} that the dimension of the conformal manifold is $6g-6=12$. There are also no relevant operators, reinforcing the picture of the $\cN=1$ MN theories as the inevitable endpoint of RG flows from the theories with higher central charge. The counting of marginal operators was done in \cite{Benini:2009mz}. For completeness we present the remaining quivers with $p=2$ in Figure \ref{otherp2}. 

\begin{figure}[ht]
\centering
\includegraphics[scale=.85]{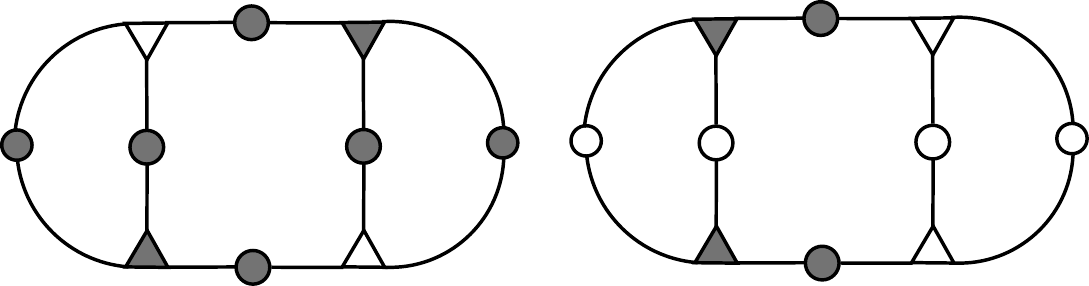}\caption{\it Two quivers with $g=3$ and $p=2$.}\label{bothp2}
\end{figure}

\begin{figure}[ht]
\centering
\includegraphics[scale=.5]{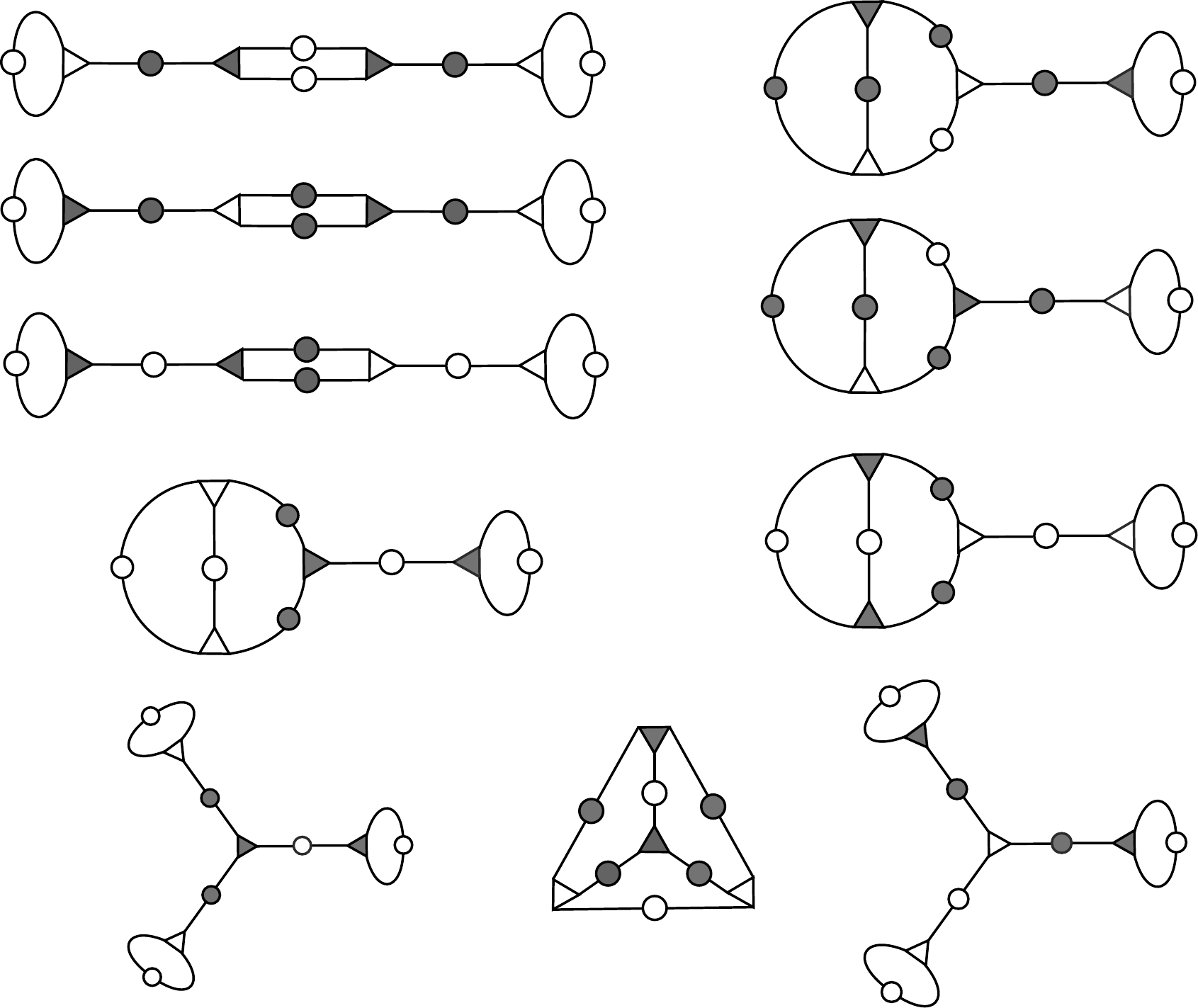}\caption{\it The remaining quivers with $g=3$ and $p=2$.}\label{otherp2}
\end{figure}
\end{itemize}

\section{Future directions}
\label{sec:outro}

In this work, we uncovered new conformal phases in the infrared dynamics of M5-branes wrapped on a complex curve $\cC_g$ in a Calabi-Yau three-fold. The main result of our construction is that for any choice of $\cC_g$ there is a one-parameter family of four-dimensional $\cN=1$ SCFTs  determined by a choice of decomposable $\bC^2$ bundle over the curve. For each such choice, we found $AdS_5$ M-theory backgrounds which describe the backreaction of the wrapped branes. Furthermore, for the theories with $g>1$ and $|z|<1$, we identified four-dimensional generalized quiver gauge theories constructed out of $T_N$ building blocks which flow to the fixed points in the infrared. 

With the introduction of such a large family of $\cN=1$ SCFTs, there are several avenues open to be explored, many of which are likely amenable to analysis by existing techniques. 
For example, the biggest question we are left with is what to make of the CFT duals of the backgrounds with $|z| > 1$ and $g>1$. It is natural to suggest that they somehow employ $T_N$'s as building blocks, but it is not yet clear how precisely to construct them. However, even the theories we find with $|z|<1$ lead to many open questions.
For example, it would be interesting to investigate the moduli space of supersymmetric vacua of these theories. This space should be realized geometrically in terms of branched coverings of the curve $\cC_g$, much as in \cite{Gaiotto:2009we}, but with the branches valued in the line bundles $\cL_i$. It should then be possible to formulate an $\cN=1$ curve describing the infrared dynamics on this moduli space \cite{Intriligator:1994sm}. 

The quiver constructions of Section \ref{sec:four} also suggest the existence of a rich web of dualities which relate different UV quivers that flow to the same SCFT in the infrared. To explore these relations, it will be necessary to make further progress on understanding the chiral rings of these theories. This lack of understanding is likely the source of some of the puzzles in Section \ref{sec:example}. A good starting point would be the case of theories with $N=2$, where at least the UV construction is weakly coupled. In addition, certain observables such as the superconformal index and the $S^4$ partition function of these theories might be computable, extending the deep structures known to exist for the $\cN=2$ theories.

One obvious shortcoming of the present work is the assumption that the curve $\cC_g$ is closed. In the case of the $\cN=2$ MN twist, there is a rich structure arising from the possible punctures of the Riemann surface, and there should surely be an analogous story for these general $\cN=1$ constructions. It may be that with an appropriate understanding of punctured curves, one could construct the $|z|>1$ theories out of a larger set of elementary building blocks, bypassing the restrictions which limited us to only studying the intermediate fixed points by field-theoretic means.

In the dual supergravity context, the inclusion of punctures was accomplished in \cite{Gaiotto:2009gz} by introducing appropriate source terms into the differential equations governing the general $\cN=2$ solutions of \cite{Lin:2004nb}. An extension of this type of analysis to introduce punctures in our $\cN=1$ backgrounds may be possible within the context of the canonical parameterization of $\cN=1$ $AdS_5$ solutions utilized in Section \ref{subsec:canonical}.

A particularly mysterious subset of the fixed points we have found are those where $\cC_g$ is a two-sphere or a torus. In these two cases, there is unlikely to be any ``building block'' approach to constructing the theories, and any field theory method for accessing their properties would be very interesting. The sphere also has the only violation of the bound of \cite{Hofman:2008ar}; the source of this violation remains mysterious.

We have specifically focused on quivers built out of $SU(N)$ vector multiplets and $SU(N)$ $T_N$ theories, \ie, theories describing the infrared dynamics of the $A_N$ $(2,0)$ theory on $\cC_g$. It is clear that these constructions should be generalizable to quivers built out of $SO(2N)$ and $Sp(2N-2)$ vector multiplets and the $SO(2N)$ $T_N$ theories to describe the infrared limit of the $D_N$ theory compactified on $\cC_g$ with a more general partial twist \cite{Tachikawa:2009rb}.

Finally, we observe that the basic philosophy which led to the discovery of the new fixed points in this paper may be applicable to the case of M5-branes wrapping three-manifolds. In particular, while the geometry near a calibrated three-cycle in a Calabi-Yau threefold will always locally take the form of the cotangent bundle, for the case of an associative three-cycle in a $G_2$-holonomy manifold there is again the freedom to choose an arbitrary $SU(2)$ bundle over the cycle. For a trivial flat $SU(2)$ bundle, there are known $AdS_4$ duals to the IR fixed point theory \cite{Gauntlett:2003di}. It may be that for appropriate choices of nontrivial bundles one can establish, at least holographically, the existence of an analogous landscape of $\cN=1$ SCFTs in three dimensions.

\begin{center}
{\bf Acknowledgements}
\end{center}
\medskip

We are grateful to Francesco Benini for numerous enlightening conversations. We would also like to thank Tudor Dimofte, Mike Douglas, Abhijit Gadde, Jerome Gauntlett, Nick Hamalgyi, James Liu, Juan Maldacena, Luca Mazzucato, Dave Morrison, Takuya Okuda, Leo Pando-Zayas, Joe Polchinski, Leonardo Rastelli, Phil Szepietowski, AJ Tolland, Balt van Rees, Dan Waldram, and Edward Witten for helpful and informative discussions. We also thank the Simons Workshop in Mathematics and Physics 2011 for a stimulating working environment during the initial stages of this project. The work of CB and NB is supported in part by DOE grant DE-FG02-92ER-40697. IB is supported by DOE grant DE-FG02-95ER-40899. BW is supported by the Fundamental Laws Initiative of the Center for the Fundamental Laws of Nature, Harvard University, and the STFC Standard Grant ST/J000469/1 ``String Theory, Gauge Theory and Duality."

\appendix
\section{BPS equations}\label{app:BPSeqns}
\renewcommand{\theequation}{A.\arabic{equation}}
\setcounter{equation}{0} 

In Section \ref{sec:seven} we pointed out that the holographic duals to the field theories of interest lie within a simple truncation of the maximal gauged supergravity in seven dimensions to the metric, two Abelian gauge fields, and two real scalars.\footnote{There is also a three-form gauge potential in this truncation, however it vanishes identically for all solutions discussed here.} This appendix contains some details related to the derivation of the BPS flow equations in this truncation.

The supersymmetry variations for the fermionic fields of the truncation are given by \cite{Liu:1999ai,Pernici:1984xx}
\bea
\delta\psi_{\mu} &=& \left[ \nabla_{\mu} +m(A_{\mu}^{(1)}\Gamma^{12} + A_{\mu}^{(2)}\Gamma^{34}) + \ds\frac{m}{4} e^{-4(\lambda_1+\lambda_2)}\gamma_{\mu} + \ds\frac{\gamma_{\mu}}{2} \gamma^{\nu} \partial_{\nu} (\lambda_1+\lambda_2) \right] \varepsilon \notag\\
&& \qquad\qquad+ \ds\frac{\gamma^{\nu}}{2} \left( e^{-2\lambda_1}F_{\mu\nu}^{(1)}\Gamma^{12} + e^{-2\lambda_2}F_{\mu\nu}^{(2)}\Gamma^{34} \right) \varepsilon~, \notag\\
\delta\chi^{(1)} &=& \left[ \ds\frac{m}{4} (e^{2\lambda_1}- e^{-4 (\lambda_1+\lambda_2)}) - \ds\frac{\gamma^{\mu}}{4}  \partial_{\mu} (3\lambda_1+2\lambda_2) - \ds\frac{\gamma^{\mu\nu}}{8}e^{-2\lambda_1}F_{\mu\nu}^{(1)}\Gamma^{12} \right] \varepsilon ~, \label{AAEq3}\\
\delta\chi^{(2)} &=& \left[ \ds\frac{m}{4} (e^{2\lambda_2}- e^{-4 (\lambda_1+\lambda_2)}) - \ds\frac{\gamma^{\mu}}{4}  \partial_{\mu} (2\lambda_1+3\lambda_2) - \ds\frac{\gamma^{\mu\nu}}{8}e^{-2\lambda_2}F_{\mu\nu}^{(2)}\Gamma^{34} \right] \varepsilon ~, \notag
\eea
where $m$ is the gauge coupling of the gauged supergravity and is related to the radius of the $AdS_7$ vacuum of the theory. Our goal is to find equations for the scalars, vectors, and metric functions in the Ansatz \eqref{M5metricAnsatz}-\eqref{M5gaugescalarAnsatz} which guarantee the existence of some spinor $\varepsilon$ for which the above variations vanish. The twist of the boundary field theory suggests the following decomposition of the seven-dimensional spinor,
\be
\gamma_{\hat{x}_1\hat{x}_2}\varepsilon=i\alpha\varepsilon~,\qquad \Gamma^{12}\varepsilon=i\beta_1\varepsilon~,\qquad \Gamma^{34}\varepsilon=i\beta_2\varepsilon~,\qquad \gamma_{\hat r}\varepsilon=\eta\varepsilon~,
\label{spinorcharges}
\ee
with $\alpha$, $\beta_1$, $\beta_2$, $\eta=\pm1$.\footnote{The symplectic Majorana spinor $\varepsilon$ transforms in the $\bf{4}$ of $SO(5)$. The $\Gamma^i$ are $SO(5)$ gamma matrices and $\gamma_{\mu}$ are seven-dimensional space-time gamma matrices. We use the notation $\gamma_{\mu_1\ldots\mu_p} = \gamma_{[\mu_1}\ldots\gamma_{\mu_p]}$ and suppress all spinor indices. Hats indicate tangent space indices.} The supersymmetries preserved by the flow should be those which restrict to Poincar\'e supersymmetries on the boundary at $r\to0_+$, which fixes $\eta=1$. Additionally, four-dimensional Poincar\'e invariance of the backgrounds implies that the spinors are constant in the $\mathbb{R}^{1,3}$ directions,
\be
\partial_{t}\varepsilon=\partial_{z_i}\varepsilon=0~.\label{AAEq8}
\ee

The conditions for the supersymmetry variations \eqref{AAEq3} to vanish are of two types. Vanishing of the variation of the dilatinos $\chi^{(i)}$ and the $(t,z_1,z_2,z_3)$ components of the gravitino $\psi_{\mu}$ imposes explicit conditions on the background fields. Alternatively, vanishing of the variations of the $(r,x_1,x_2)$ components of the gravitino imply that the spinor  solves a certain system of partial differential equations. Integrability of this system imposes additional constraints on the background fields. After a straightforward calculation, the following system of BPS equations emerges:\footnote{We have set $\alpha=\beta_1=\beta_2=1$. One can check that only one other choice of signs yields the same equations, and consequently the generic solution of these equations will preserve one quarter of the maximal supersymmetry.}
\bea
&&  \partial_{r}(3\lambda_1+2\lambda_2) - m e^{h+2\lambda_1} +m e^{h-4\lambda_1-4\lambda_2} - e^{h - 2 \hat{g} - 2\lambda_1} F^{(1)}_{x_1x_2}=0~, \label{BPSeqn1}\\
&&  \partial_{r}(2\lambda_1+3\lambda_2) - m e^{h+2\lambda_2} +m e^{h-4\lambda_1-4\lambda_2} - e^{h - 2 \hat{g} - 2\lambda_2} F^{(2)}_{x_1x_2}=0~, \label{BPSeqn2}\\
&& (\partial_{x_1}+i\partial_{x_2})(3\lambda_1+2\lambda_2) - e^{-h - 2\lambda_1 }(F^{(1)}_{x_2r}-iF^{(1)}_{x_1r})=0~, \label{BPSeqn3}\\
&& (\partial_{x_1}+i\partial_{x_2})(2\lambda_1+3\lambda_2) - e^{-h - 2\lambda_2 }(F^{(2)}_{x_2r}-iF^{(2)}_{x_1r})=0~, \label{BPSeqn4}\\
&& \partial_{r}\left(f +\lambda_1+\lambda_2\right) + \ds\frac{m}{2} e^{h-4\lambda_1-4\lambda_2} = 0~, \label{BPSeqn5}\\
&& \partial_{x_1}\left(f +\lambda_1+\lambda_2\right) = \partial_{x_2}\left(f +\lambda_1+\lambda_2\right) = 0~, \label{BPSeqn6}\\
&& \partial_{r}(\hat{g} - 4\lambda_1 - 4\lambda_2) +  m e^{h +2\lambda_1} +  m e^{h +2\lambda_2} - \ds\frac{3m}{2} e^{h - 4\lambda_1 - 4\lambda_2} =0 ~, \label{BPSeqn7}\\
&& \partial_{r}\partial_{x_2}(\hat{g}-4\lambda_1-4\lambda_2) + 2m F^{(1)}_{rx_1}+ 2m F^{(2)}_{rx_1} = 0~, \label{BPSeqn8}\\
&& \partial_{r}\partial_{x_1}(\hat{g}-4\lambda_1-4\lambda_2) + 2m F^{(1)}_{x_2r}+ 2m F^{(2)}_{x_2r} = 0~, \label{BPSeqn9}\\
&& (\partial_{x_1}^2 + \partial_{x_2}^2)(\hat{g}-4\lambda_1-4\lambda_2) - 2m F^{(1)}_{x_1x_2} - 2m F^{(2)}_{x_1x_2}=0~. \label{BPSeqn10} 
\eea
These equations are valid with no assumptions on the form of the function $\hat{g}(x_1,x_2,r)$. Upon restricting to the case of a constant curvature metric on $\cC_g$ at fixed $r$, there is a dramatic simplification of the equations.\footnote{From this point forward  we adopt the conventions of \cite{Maldacena:2000mw} and fix the normalization $m=2$. We further assume $\cC_g$ is of genus $g\neq1$, delaying the analysis of the torus until Appendix \ref{app:T2}.} The simplifying assumption is that 
\begin{align}
\begin{split}
&\hat{g}(x_1,x_2,r) = g(r) - \log x_2\;\;\qquad\qquad\qquad~, \qquad \kappa=-1~, \\
&\hat{g}(x_1,x_2,r) = g(r) - 2\log\left(\ds\frac{1+x_1^2+x_2^2}{2}\right)~, \qquad \kappa=+1 ~,
\end{split} 
\end{align}
and consequently the BPS equations reduce to the following system of ODEs:
\begin{align}
\begin{split}
& f' + \lambda_1'+\lambda_2'+ e^{f-4(\lambda_1+\lambda_2)}=0~, \\
& g' + \lambda_1'+\lambda_2' + \ds\frac{\kappa+\kappa z}{8} e^{f-2g-2\lambda_1}+ \ds\frac{\kappa-\kappa z}{8} e^{f-2g-2\lambda_2}+ e^{f-4(\lambda_1+\lambda_2)}=0~, \\
& 3\lambda_1'+2\lambda_2' + 2e^{f-4(\lambda_1+\lambda_2)} - 2e^{f+2\lambda_1} +  \ds\frac{\kappa+\kappa z}{8} e^{f-2g-2\lambda_1} =0~, \\
& 2\lambda_1'+3\lambda_2' + 2e^{f-4(\lambda_1+\lambda_2)} - 2e^{f+2\lambda_2} + \ds\frac{\kappa-\kappa z}{8} e^{f-2g-2\lambda_2} =0~,
\label{BPSeqns}
\end{split}
\end{align}
where a prime denotes differentiation with respect to $r$. 

Although we do not explore this in the current work, we would like to comment on the possibility of studying the BPS flows for a general choice of metric on $\cC_g$. As was discussed in \cite{Anderson:2011cz}, the relevant twists of the $(2,0)$ theory can be implemented for any metric on the curve. This is easily seen from the BPS equations above, which admit asymptotically locally $AdS_7$ solutions with an arbitrary choice for the warp factor $\hat{g}(x_1,x_2)$ in the UV. It was shown in \cite{Anderson:2011cz} that the holographic RG flow for the two MN theories uniformizes the metric on $\cC_g$, and in the IR all solutions limit to $AdS_5$ fixed points with constant curvature metric. It seems likely that this analysis could be extended to the more general holographic RG flows discussed in Section \ref{subsec: holoRGflows}. Indeed, a linearized UV and IR analysis indicates that the conclusions of \cite{Anderson:2011cz} apply to the theories discussed in the current work as well. In particular, this means that the choice of complex structure is the only metric information which is not ``washed out" by the RG flow, remaining as marginal deformations of the four-dimensional SCFT.

\section{Equations of motion}\label{app:eom}
\renewcommand{\theequation}{B.\arabic{equation}}
\setcounter{equation}{0} 

There is a subtlety in the discussion above. Namely, it is not always true that a given solution to the supergravity BPS equations is a solution to the equations of motion. Here we explain that for the Ansatz and truncation in question, the BPS equations {\it do} imply the equations of motion.

The equations of motion for the fields in the truncation of interest were derived in \cite{Liu:1999ai}. We further impose that the three-form gauge potential vanishes; this is consistent with all equations of motion for the Ansatz \eqr{M5metricAnsatz}-\eqr{M5gaugescalarAnsatz}. The Maxwell equations are
\be
\nabla^{\mu}(e^{-4\lambda_1} F^{(1)}_{\mu\nu}) = 0~, \qquad\qquad \nabla^{\mu}(e^{-4\lambda_2} F^{(2)}_{\mu\nu}) = 0~, 
\label{Maxeqns}
\ee
and the scalar equations are 
\begin{align}
\begin{split}
\nabla^2(3\lambda_1+2\lambda_2) +e^{-4\lambda_1}F^{(1)}_{\mu\nu} F^{(1)\mu\nu} -\ds\frac{m^2}{8} \partial_{\lambda_1}V =0 ~,\\[4pt]
\nabla^2(2\lambda_1+3\lambda_2) +e^{-4\lambda_2}F^{(2)}_{\mu\nu} F^{(2)\mu\nu} -\ds\frac{m^2}{8} \partial_{\lambda_2}V = 0~,
\label{scalarEoM}
\end{split}
\end{align}
where the potential is given by
\be
V = -8 e^{2\lambda_1+2\lambda_2} -4 e^{-2\lambda_1-4\lambda_2} -4 e^{-4\lambda_1-2\lambda_2} +e^{-8\lambda_1-8\lambda_2}~.
\ee
Subject to the Ansatz, the non-vanishing components of Maxwell's equations are
\bea
&&\nabla^{\mu}(e^{-4\lambda_i} F^{(i)}_{\mu r}) = \partial_{x_1}F^{(i)}_{rx_1} +F^{(i)}_{rx_1}\partial_{x_1}( 4f - h - 4\lambda_i) + \partial_{x_2}F^{(i)}_{rx_2} +F^{(i)}_{rx_2}\partial_{x_2}( 4f - h - 4\lambda_i)~, \notag\\[4pt]
&&\nabla^{\mu}(e^{-4\lambda_i} F^{(i)}_{\mu x_1}) =e^{2\hat{g}}[ \partial_{r}F^{(i)}_{rx_1} +F^{(i)}_{rx_1}\partial_{r}( 4f - h - 4\lambda_i) ]\notag \\
&& \qquad\qquad\qquad\qquad\qquad - e^{2h}[ \partial_{x_2}(x_2^2F^{(i)}_{x_1x_2}) +x_2^2F^{(i)}_{x_1x_2}\partial_{x_2}( 4f + h - 2g - 4\lambda_i) ]~, \\[4pt]
&&\nabla^{\mu}(e^{-4\lambda_i} F^{(i)}_{\mu x_2}) =e^{2\hat{g}}[ \partial_{r}F^{(i)}_{rx_2} +F^{(i)}_{rx_2}\partial_{r}( 4f - h - 4\lambda_i) ]\notag \\
&& \qquad\qquad\qquad\qquad\qquad + e^{2h}[ \partial_{x_1}(x_2^2F^{(i)}_{x_1x_2}) +x_2^2F^{(i)}_{x_1x_2}\partial_{x_1}( 4f + h - 2g - 4\lambda_i) ]~.\notag
\eea
After a somewhat tedious calculation one can show that the BPS equations \eqref{BPSeqn1}-\eqref{BPSeqn6} imply that the Maxwell equations are satisfied. 
Using equations \eqref{BPSeqn1}-\eqref{BPSeqn10} one can also show that the two scalar equations of motion reduce to
\be
\partial_{[r}F^{(1)}_{x_1x_2]}=0~,\qquad\qquad
\partial_{[r}F^{(2)}_{x_1x_2]}=0~,
\label{Bianchi}
\ee
where the square brackets denote antisymmetrization with respect to all indices. These equations are precisely the Bianchi identities for the two gauge fields. We therefore conclude that the system of BPS equations along with Bianchi identities imply that the equations of motion are satisfied.\footnote{There is a caveat here. We have assumed that integrability of the supersymmetry variations  ensures that the Einstein equations are satisfied. This has been shown for other supergravity theories in \cite{Gauntlett:2002nw,Gauntlett:2002fz,Gauntlett:2005ww}, but has not been shown for the maximal gauged supergravity in seven dimensions. We have checked that all the $AdS_5$ backgrounds described in this paper do indeed solve the Einstein equations.} 

The Bianchi identity for the linear combination of field strengths $F^{(1)}+F^{(2)}$ actually follows directly from the system of BPS equations (more precisely from equations \eqref{BPSeqn8}-\eqref{BPSeqn10}). Alternatively, one has to impose the Bianchi identity for $F^{(1)}-F^{(2)}$ by hand.\footnote{The solutions studied in \cite{Anderson:2011cz} have either $F^{(2)}=0$ or $F^{(1)}=F^{(2)}$, so there was no need to impose a Bianchi identity by hand.} We have checked that this Bianchi identity is indeed satisfied for the solutions studied in this paper.

\section{M5-branes on the torus}\label{app:T2}
\renewcommand{\theequation}{C.\arabic{equation}}
\setcounter{equation}{0} 

The analysis of Sections \ref{sec:six} and \ref{sec:seven} focused on the case that the curve $\cC_g$ is {\it not} a torus. Here we explore the infrared dynamics of wrapped M5-branes in this special case. Since the torus admits a flat metric, we can wrap M5-branes on it and still preserve the maximal amount of supersymmetry by fixing the normal bundle to be trivial. The four-dimensional IR theory is then $\mathcal{N}=4$ SYM. In the eleven-dimensional supergravity background describing the backreaction of this brane setup, the tension of the M5-branes shrinks the volume of the torus and the background becomes singular. This singular behavior is responsible for the $N^2$ scaling of the central charge of $\mathcal{N}=4$ SYM and indicates that eleven-dimensional supergravity is not suitable for describing the gravity dual of $\mathcal{N}=4$ SYM. Instead, one should dualize to the usual type IIB description.

However, there is a more general construction with M5-branes compactified on the two-torus. One can configure $T^2$ to have two nontrivial line-bundles of equal and opposite degree fibered over it. This corresponds to taking $p=-q$ in \eqref{chernnumbers}, and preserves only a quarter of the maximal supersymmetry.

The central charges of an IR fixed point of this compactification can by computed by an anomaly calculation analogous to the one in Section \ref{subsec:cc6}. For a flat torus the tangent bundle is trivial and the choice of normal bundle and nontrivial R-symmetry bundle is encoded in the following choice of Chern roots,
\be
\cL_1 \to \cL^{\otimes z} + (1+\epsilon) c_1(F)~, \qquad\qquad  \cL_2 \to \cL^{-\otimes z}+ (1-\epsilon) c_1(F)~,
\ee
where $z\in \mathbb{Z}$ and $\cL$ is a line bundle of degree one over $T^2$. To find the central charges, we integrate the anomaly eight-form $I_{8}[G]$ over $T^2$, identify the result with the four-dimensional anomaly six-form $I_6$, and determine the parameter $\epsilon$ by $a$-maximization. The result is that
\be
\epsilon=-\ds\frac{1}{3}\ds\sqrt{\ds\frac{1+3\eta}{1+\eta}}~,
\ee
and the central charges are given by
\be
a = \ds\frac{|z|}{48} \ds\frac{r_G (1+3\eta)^{3/2}}{\sqrt{1+\eta}}~,\qquad\qquad
c = \ds\frac{|z|}{48} \ds\frac{r_G (2+3\eta) \sqrt{1+3\eta}}{\sqrt{1+\eta}}~,
\ee
where $\eta$ is defined in \eqref{etazetadef}. For the $A_N$ theory in the large $N$ limit, this yields
\be
a = c \approx \ds\frac{\sqrt{3}}{16} |z|N^3 ~.
\label{acT2anom}
\ee

To study the backreaction of this brane setup and the corresponding IR fixed point one has to solve the BPS equations \eqref{BPSeqn1}-\eqref{BPSeqn10} and find $AdS_5\times T^2$ solutions and holographic RG flows to them.  To do this we take the metric functions and background fields to have the following form,
\bea
&&\hat{g} = g_0~, \qquad\qquad\qquad\qquad h=f=f_0 - \log r~,\notag \\
&&F^{(1)}_{x_1x_2} = \ds\frac{z}{8} ~,\qquad\qquad\qquad~ F^{(2)}_{x_1x_2} = -\ds\frac{z}{8}~,\\
&&F^{(i)}_{rx_1} = F^{(i)}_{rx_2} = 0~,\qquad\quad\;\;\;  \lambda_{i}=\text{const}~,\notag
\eea
where $g_0$ and $f_0$ are constants and $z\in \mathbb{Z}$ as above. The BPS equations \eqref{BPSeqn1}-\eqref{BPSeqn10} then lead to the following one parameter family of solutions (note that we  have fixed $m=2$)
\begin{align}
\begin{split}
e^{2\lambda_1} &= \left(\ds\frac{33+19\sqrt{3}}{4}\right)^{1/5} ~, \qquad\qquad e^{2\lambda_2} = \left(\ds\frac{33-19\sqrt{3}}{4}\right)^{1/5} ~,\\
e^{2g_0} &=  \ds\frac{3^{3/10}}{2^{2/5}} \ds\frac{|z|}{8}~, \qquad\qquad\qquad\qquad\, e^{f_0} = e^{4\lambda_1+4\lambda_2} =\ds\frac{3^{2/5}}{2^{6/5}} ~.
\end{split}
\end{align}
There are holographic BPS flows for each value of $z$ from an asymptotically locally $AdS_7$ region with $\mathbb{R}^{1,3}\times T^2$ boundary to these $AdS_5\times T^2$ fixed points in the IR. These solutions are analogous to the ones presented in Section \ref{subsec: holoRGflows} for the case of twisted $T^2$ compactification.

Using the results of \cite{Cvetic:1999xp}, these solutions can be uplifted to eleven dimensions. The resulting background is similar to the one in Section \ref{subsec:trivialuplift}, so we do not present it here explicitly. We can use the eleven-dimensional background to compute the central charges of the dual field theories at large $N$. The result is
\be\label{ccT2grav}
a=c = \ds\frac{8}{3}  e^{2g_0+3h_0} = \ds\frac{\sqrt{3}}{16} |z| N^3~,
\ee
which matches the anomaly calculation \eqref{acT2anom}. 

The dimension of the operator dual to an M2 brane wrapped on the torus is also computable from the supergravity solution, and one finds
\be
\Delta(\cO_{M2}) = 4 e^{f_0+2g_0-2(\lambda_1+\lambda_2)} N = \ds\frac{\sqrt{3}}{2} |z|N~.
\ee
These $AdS_5\times T^2$ quarter-BPS solutions are similar in spirit to the supersymmetric $AdS_3\times T^2$ solutions studied recently in \cite{Almuhairi:2011ws,Donos:2011pn}.

\section{Supergravity solutions in canonical coordinates}\label{a:GMSW}
\renewcommand{\theequation}{D.\arabic{equation}}
\setcounter{equation}{0} 

In Section \ref{subsec:canonical} we presented solutions of eleven-dimensional supergravity in the canonical conventions of \cite{Gauntlett:2004zh}. This appendix describes the detailed derivation of those solutions. The starting point is a metric Ansatz of the form
\bea
ds^2_{11} &=& e^{2\lambda(y,z)} \left[ ds^2_{AdS_5} + e^{2\nu +2A(x)} (dx_1^2+dx_2^2)\right] \\
&+& e^{-4\lambda(y,z)}\left[ e^{2B(y,z)} dz^2 + e^{2C(y,z)}(d\chi + V)^2 + \sec^2\zeta dy^2 + \ds\frac{e^{6\lambda(y,z)\cos^2\zeta}}{9}(d\psi+\rho) \right]~. \notag
\eea
We have denoted the unit radius metric on $AdS_5$ by $ds^2_{AdS_5}$ and the function $A(x)$ is the conformal factor of the constant curvature metric on the Riemann surface $\Sigma_g$, obeying
\be
(\partial_{x_1}^2 + \partial_{x_2}^2) A + \kappa e^{2A} = 0~. \label{Adef}
\ee
The constant $\kappa$ is the Gaussian curvature of the Riemann surface and we set $\kappa=1$ for $S^2$, $\kappa=-1$ for a hyperbolic surface and $\kappa=0$ for the torus.\footnote{In most of the expressions that follow, the limit $\kappa \to 0$ should be taken with some care. We will comment on this at the end of this appendix.} The one-form $V$ is defined such that 
\be
dV = \ds\frac{\kappa}{2-2g} e^{2A} dx_1 \wedge dx_2~, \qquad\qquad \int_{\mathcal{C}} dV = 2\pi~.
\ee
In addition, we introduce the real constant $\nu$, which will parametrize the solutions.

An important ingredient in the analysis of \cite{Gauntlett:2004zh} is the realization of the internal six-manifold locally as a fibration over a four-dimensional K\"ahler manifold. In the Ansatz above, this manifold is a two-dimensional fibration over the curve $\cC_g$
\be
ds^2_{4} =  e^{6\lambda + 2\nu+2A} (dx_1^2 + dx_2^2) +e^{2B} dz^2 +e^{2C} (d\chi+V)^2 ~,
\label{KEbase}
\ee
and it will be convenient to define the complex vielbein 
\be
\mathrm{e}_1 = e^{3\lambda + \nu + A}(dx_1+ i dx_2)~, \qquad\qquad \mathrm{e}_2  = e^{B}dz + i e^{C} (d\chi+ V)~.
\ee
The K\"ahler form and the holomorphic $(2,0)$ form on this space are denoted by $\hat{J}$ and $\hat{\Omega}$, and read
\be
\hat{J} = \ds\frac{i}{2} (\mathrm{e}_1 \wedge \bar{\mathrm{e}}_1 + \mathrm{e}_2\wedge \bar{\mathrm{e}}_2)~, \qquad\qquad \hat{\Omega} = e^{i\psi} \mathrm{e}_1\wedge \mathrm{e}_2~.
\label{Kahlerform}
\ee
The conditions for the existence of a Killing spinor in eleven-dimensional supergravity within this Ansatz were derived in \cite{Gauntlett:2004zh} and can be written as
\bea
&& d_{4}\hat{J} = 0~, \label{eqn1}\\
&& \partial_{y}\hat{J} = - \ds\frac{2}{3} y d_{4}\rho~, \label{eqn2}\\
&& \partial_{\psi}\hat{J} = 0~, \label{eqn7} \\
&&  e^{6\lambda}\sin^2\zeta = 4 y^2~, \label{eqn3}\\
&& d_{4}\hat{\Omega} = (i\rho - d_{4}\log\cos\zeta)\wedge \hat{\Omega}~, \label{eqn4}\\
&& \partial_{y} \hat{\Omega} = \left( -\frac{3\tan^2\zeta}{2y} - \partial_{y} \log\cos\zeta \right) \hat{\Omega}~, \label{eqn5}\\
&& \partial_{\psi}\hat{\Omega} = i \hat{\Omega}~,\label{eqn6}
\eea
where the exterior derivative on the K\"ahler manifold is denoted by $d_4$. Note that none of the background functions depend on $\psi$ and $\chi$, which are the two isometries of the compact manifold. The construction of \cite{Gauntlett:2004zh} guarantees that the Killing vector $\partial_{\psi}$ corresponds to the superconformal R-symmetry of the dual $\mathcal{N}=1$ SCFT.

It is clear from the explicit expressions for $\hat{J}$ and $\hat{\Omega}$ in \eqref{Kahlerform} that equations \eqref{eqn7} and \eqref{eqn6} are trivially satisfied. Equation \eqref{eqn1} implies the following differential constraint,
\be
\partial_{z}e^{6\lambda} = \ds\frac{\kappa}{2-2g} e^{B+C-2\nu}~.\label{EQN1}
\ee
The constraint coming from \eqref{eqn5} can be written as
\be
\partial_{y}[B(y,z)-C(y,z)] = 0~.
\ee
This is solved by imposing $B(y,z) = C(y,z) + H(z)$, where $H(z)$ is an unfixed function of $z$ which can be set to zero by a reparametrization of the $z$-coordinate.
In terms of the function
\be
e^{2\gamma} = e^{6\lambda}\cos^2\zeta = e^{6\lambda} - 4y^2~,
\ee
\eqref{eqn4} implies that the one-form $\rho$ is given by
\be
\rho  = \partial_{x_2}A dx_1 - \partial_{x_1}A dx_2 - \partial_{z}(\gamma+C) (d\chi+V)~.
\ee
The constraints coming from \eqref{eqn2} can be written as
\begin{align}
\begin{split}
& \partial_{y}e^{6\lambda} = -\ds\frac{2\kappa ye^{-2\nu}}{3}\left(1-\ds\frac{1}{(2-2g)} \partial_{z}(\gamma+C)\right)~,\\
& \partial_{y}e^{2C} = \ds\frac{2 y}{3} \partial_{z}^2(\gamma+C)~. \label{EQN3}
\end{split}
\end{align}
Finally, the constraint in \eqref{eqn5} takes the form
\be
\partial_{y}(\gamma+C) = -6ye^{-2\gamma}~. \label{EQN4}
\ee

From equations \eqref{EQN1},  \eqref{EQN3}, and \eqref{EQN4} one derives the following coupled system of PDEs for the function $\gamma$:
\begin{align}
\begin{split}
\partial_{y}e^{4\gamma} &= \ds\frac{2y}{3} f(y) - 24y e^{2\gamma}~,  \\
\partial_{z}e^{4\gamma} &= G(y) +a_1 f(y) e^{2\gamma} +a_2 e^{4\gamma}~. \label{eqngamma}
\end{split}
\end{align}
The functions $f(y)$ and $G(y)$ arise from integrating some intermediate differential equations and are as yet undetermined. For convenience we have defined the constants $a_1$ and $a_2$ to be
\be
a_1 \equiv \ds\frac{2(2-2g)e^{2\nu}}{\kappa }~, \qquad\qquad a_2 \equiv - \ds\frac{12(2-2g)e^{2\nu}}{\kappa } \left(1 - \ds\frac{\kappa}{6e^{2\nu}}\right)~.
\ee
The functions $f(y)$ and $G(y)$ can be determined by imposing integrability for the system \eqref{eqngamma}. The result is\footnote{The constant term in the function $f$ is an integration constant. One can show that it parametrizes the freedom to rescale the eleven-dimensional metric by an overall positive constant. We choose to set it to $1$.}
\be
f = 1 + \ds\frac{6 a_2 }{a_1} y^2~, \qquad\qquad G = - \ds\frac{a_1}{36} f^2~.
\label{fdefinition}
\ee
To proceed further it is convenient to introduce a new coordinate $q$ defined by
\be
e^{2\gamma} = q f(y)~,
\ee
in terms of which we have the following relation on one-forms,
\be
dz = \ds\frac{2q}{a_1 k(q)} dq + \ds\frac{24 y}{f(y)a_1} dy~,
\ee
where we have defined
\be
k(q) = \ds\frac{a_2}{a_1} q^2 + q - \ds\frac{1}{36}~.
\ee
It is useful to note that in the coordinates $(y,q)$ we have
\be
e^{2C} = \ds\frac{a_1^2}{4} \ds\frac{f(y)k(q)}{q}~,
\ee
and that the one-form $\rho$ can be rewritten as
\be
\rho = (2-2g) V - \ds\frac{1}{2}(a_2 + \ds\frac{a_1}{2q}) (d\chi + V)~. \label{rho}
\ee
The eleven-dimensional metric in this new coordinate system is given by
\bea
ds^2_{11} &=& e^{2\lambda(y,z)} \left[ ds^2_{AdS_5} + e^{2\nu +2A(x)} (dx_1^2+dx_2^2)\right]+  e^{-4\lambda(y,z)}ds^2_{M_4}~,
\eea
where
\begin{align}
\begin{split}
ds^2_{M_4} = \left( 1+ \ds\frac{4 y^2}{q f(y)} \right) dy^2 &+ \ds\frac{ f(y) q}{k(q)} \left(dq + \ds\frac{12  y k(q)}{f(y) q} dy\right)^2 \\[4pt]
&+ \ds\frac{ a_1^2}{4}~ \ds\frac{f(y)k(q)}{q}(d\chi +V)^2 + \ds\frac{q f(y)}{9} (d\psi + \rho)^2~.
\end{split}
\end{align}
It is clear that to have a positive definite metric we should impose the constraints
\be
qf(y) \geq 0~, \qquad\qquad k(q) \geq 0~.
\ee
From these constraints we find the ranges of the coordinates $y$ and $q$, 
\be
q_{-} \leq q \leq q_{+}~, \qquad\qquad y_{-} \leq y \leq y_{+}~, 
\ee
where
\be
q_{\pm} = - \ds\frac{a_1}{2a_2} \left( 1 \pm \ds\sqrt{1+ \ds\frac{a_2}{9a_1}} \right)~, \qquad\qquad y_{\pm} = \pm \ds\sqrt{\ds\frac{-a_1}{6a_2 }}~.
\ee
The constants $q_{\pm}$ and $y_{\pm}$ are the zeroes of the functions $k(q)$ and $f(y)$. Note that in order to have a compact range for $q$ the constant $\nu$ must satisfy the following bounds,
\be
- 9 \leq \kappa e^{-2\nu} - 6 \leq 0~. \label{nupositivity}
\ee

To compare these backgrounds with the uplifted solutions of Section \ref{subsec:trivialuplift} we want to find angular coordinates $\phi_+$ and $\phi_-$ which fiber independently over $\cC_g$ and express them in terms of $d\chi$ and $d\psi$. This is accomplished by the following change of variables,
\begin{eqnarray}
d\psi &=& \ds\frac{q_{-} (a_1+2 a_2 q_{+})}{a_1(q_{-}-q_{+})} \left( \ds\frac{36 q_{+} (a_1+36a_2q_{+}^2)}{a_2(q_{+}-q_{-})} \right)^{1/2} d\phi_{+} \notag\\
&& \qquad\qquad\qquad - \ds\frac{q_{+} (a_1+2 a_2 q_{-})}{a_1(q_{-}-q_{+})} \left( \ds\frac{36 q_{-} (a_1+36a_2q_{-}^2)}{a_2(q_{+}-q_{-})} \right)^{1/2} d\phi_{-}~, \\
d\chi &=& \ds\frac{4q_{+}q_{-}}{a_1(q_{-}-q_{+})} \left[ \left( \ds\frac{36 q_{+} (a_1+36a_2q_{+}^2)}{a_2(q_{+}-q_{-})} \right)^{1/2}d\phi_{+} - \left(\ds\frac{36 q_{-} (a_1+36a_2q_{-}^2)}{a_2(q_{+}-q_{-})} \right)^{1/2}d\phi_{-} \right]~.\notag
\label{dpsidchi}
\end{eqnarray}
The normalizations have been chosen so that the periods of the coordinates $\phi_{\pm}$ are both $2\pi$. One can then rewrite the metric as
\begin{align}
\begin{split}
ds^2_{M_4} &=   \left( 1 + \ds\frac{4 y^2}{q f(y)} \right) dy^2 + \ds\frac{ f(y) q}{k(q)} \left(dq + \ds\frac{12  y k(q)}{f(y) q} dy\right)^2 \\[8pt]
&+ \ds\frac{4 a_1 q_{+} f(y) (q-q_{+})}{a_2 (q_{+}-q_{-})}(d\phi_{+} -m_{+} V)^2 + \ds\frac{4 a_1 q_{-} f(y) (q-q_{-})}{a_2 (q_{+}-q_{-})}(d\phi_{-} -m_{-} V)^2~. 
\label{4metricnice}
\end{split}
\end{align}
Here, we have defined
\be
m_{\pm} \equiv (g-1) \left(1 \mp \ds\frac{2}{\kappa} \sqrt{3 e^{4\nu} +\kappa e^{2\nu}}\right) ~.
\ee
To have a well-defined fibration over $\cC_g$, the real numbers $m_{\pm}$ have to be integers. These integers are not independent, and in fact satisfy the same condition which was automatically imposed by the Calabi-Yau condition in the brane construction,
\be
m_{+}+m_{-} = 2(g-1)~.
\ee
Parameterizing this choice of integers by a single rational number
\be
z \equiv \ds\frac{m_{+}}{g-1}-1~, 
\ee
in terms of which
\be
m_{\pm}  = (g-1) (1\pm z)~,
\ee
we find that
\be
e^{2\nu} = \ds\frac{1}{6} ( -\kappa + \sqrt{\kappa^2 +3 \kappa^2 z^2} )~.
\ee

At this point the parameter $z$ is just a label, but in fact it is precisely the same parameter used to label the seven-dimensional $AdS_5$ fixed points in Section \ref{subsec:N=1fp} and their eleven-dimensional uplift. The integers $m_{\pm}$ are precisely the integers $p$ and $q$ defined in \eqref{chernnumbers} and the angles $\phi_{\pm}$ correspond to the angles $\phi_{1,2}$ in the supergrvaity solution of Section \ref{subsec:trivialuplift}. Note that the explicit expression for $e^{2\nu}$ guarantees that the positivity bounds in \eqref{nupositivity} are obeyed. One can check that there are no conical singularities at the ends of the $q$ and $y$ intervals where the orbits of the Killing vectors $\ell^{\mu}_{\pm}=\partial_{\phi_{\pm}}$ degenerate. To show this we follow the analysis of \cite{Cvetic:2005ft,Cvetic:2005vk} and show that the ``surface gravity", $\kappa_{\pm}$, for each degeneration locus is equal to one for the metric in \eqref{4metricnice}.

It is useful to express the coordinate transformation \eqref{dpsidchi} in terms of the twist parameter $z$,
\be
d\psi = d\phi_{+} + d \phi_{-}~, \qquad\qquad d\chi = \ds\frac{1}{1+\epsilon} d\phi_{+} - \ds\frac{1}{1-\epsilon} d\phi_{-}~.
\ee
where the parameter $\epsilon$ is the same as the one defined in \eqref{epslargeN}. This in turn leads to the following linear relation between the Killing vectors generating the isometries of the background,
\be
\partial_{\psi} = (1+\epsilon)\partial_{\phi_{+}} + (1-\epsilon) \partial_{\phi_{-}} ~, \qquad\qquad \partial_{\chi}  = \partial_{\phi_{+}}  - \partial_{\phi_{-}} ~.
\ee
The utility of the coordinates of \cite{Gauntlett:2004zh} is now manifest. The Killing vector $\partial_{\psi}$ is precisely the correct linear combination of the two isometries $\partial_{\phi_{\pm}}$ needed to ensure that it is dual to the R-symmetry of the infrared SCFT for a given choice of the parameter $z$. The result matches that of the anomaly analysis in Section \ref{sec:six} and the field theory approach of Section \ref{sec:four}.

To have a complete supergravity solution we need to specify the four-form flux. It was shown in \cite{Gauntlett:2004zh} that it can be written as 
\begin{align}
\begin{split}
G_{(4)} = - (\partial_{y}e^{-6\lambda}) \widehat{\text{vol}}_{4} &- e^{-9\lambda} \sec\zeta (\hat{*}_4d_4 e^{6\lambda})\wedge K^{1}\\
 &\qquad\qquad\qquad+\ds\frac{e^{3\lambda}}{3}(\cos^2\zeta \hat{*}_4 d_{4}\rho -12 e^{-6\lambda}\hat{J})\wedge K^{1}\wedge K^{2}~,
\end{split}
\end{align}
where
\be
K^{1} \equiv e^{-3\lambda}\sec\zeta dy~, \qquad\qquad K^{2} \equiv \ds\frac{\cos\zeta}{3}(d\psi+\rho)~.
\ee
Quantities with hats refer to the four-dimensional K\"ahler-Einstein base \eqref{KEbase}, and $\hat{J}$ is the K\"ahler from \eqref{Kahlerform}.

For completeness we compute the central charges of the dual SCFTs to these $AdS_5$ solutions of eleven-dimensional supergravity. We split the four-form flux into two parts,
\be
G_{(4)} = G_{(2)}\wedge dx_1\wedge dx_2 + G_{\perp}~,
\ee
where $G_{\perp}$ is a four-form with no legs along $dx_i$, and one can show that 
\be
G_{\perp} = \ds\frac{4}{3} \ds\frac{a_1}{a_2\sqrt{1+\frac{a_2}{9a_1}}} \ds\frac{f}{qf+4y^2}\left( 2+\ds\frac{1-12q}{12}\ds\frac{f}{qf+4y^2} \right) dq\wedge dy\wedge d\phi_{+}\wedge d\phi_{-}~.  
\ee
To get the properly normalized central charge we use the conventions of Section 2 of \cite{Gauntlett:2006ai}. The number of M5-branes is obtained by integrating the flux component $G_{\perp}$ over the topological $S^4$ transverse to the Riemann surface,
\be
N = \ds\frac{1}{(2\pi)^3 \ell_{11}^3} \int_{S^4} G_{\perp}~,
\ee
where $\ell_{11}^3$ is the eleven-dimensional Planck length. It is useful to define
\be
I (\kappa,z) \equiv \int_{S^4} G_{\perp}=\ds\frac{8\pi^2}{9}\left( \ds\frac{\kappa-3\kappa z^2 -\sqrt{\kappa^2(1+3z^2)}}{3\kappa-3\kappa z^2} \right)^{3/2}~.
\ee
The central charge of the dual field theory is then given by \cite{Gauntlett:2006ai}
\begin{equation}
c = \ds\frac{1}{2^7 \pi^6 \ell_{11}^9} \ds\frac{2e^{2\nu}a_1}{3 a_2 \sqrt{1+\frac{a_2}{9a_1}}}\int f(y) e^{2A} dx_1\, dx_2 \, dy\,dq\, d\phi_{+}\, d\phi_{-}~, \label{GMSWccgen}
\end{equation}
and a short calculation yields
\be
c = \ds\frac{2^7 \pi^6 (g-1)}{3\kappa} \ds\frac{e^{2\nu} (y_{+}-y_{-})^3 (q_{+}-q_{-})}{I(\kappa,z)^3 \sqrt{1+\frac{a_2}{9a_1}}} N^3~.
\ee
For $\kappa=\pm1$, \ie, for a Riemann surface with $g\neq1$, this reduces to
\be
c = \ds\frac{(1-g) N^3}{\kappa}\left( \ds\frac{\kappa-\kappa9z^2+(\kappa^2+3z^2)^{3/2}}{48z^2}\right)~,
\ee
which matches the central charge computed in \eqref{ccuplift}.  Note that for $g=1$, taking the limit $\kappa \to 0$ with $\kappa/(g-1)\to1$ and $\kappa z \to z \in \mathbb{Z}$ in \eqref{GMSWccgen} reproduces the central charges for the $AdS_5\times T^2$ solutions of \eqref{ccT2grav}.

\section{SCFT miscellany}\label{a:scftrev}
\renewcommand{\theequation}{E.\arabic{equation}}
\setcounter{equation}{0} 

An $\cN=2$ SCFT has R-symmetry $SU(2)_R \times U(1)_R$. A convenient basis for the Cartan subalgebra of this R-symmetry is $I_3$, generated by ${\rm diag}(1,-1)  \subset SU(2)_R$, and the generator of $U(1)_R$, denoted $R_{\mathcal{N}=2}$. For free vector and chiral superfields, the charge assignments of these fields are given in the following table.
\be
 \begin{array}{c|ccc}  R_{\mathcal{N}=2} \setminus I_3 & \frac{1}{2} & 0 & -\frac{1}{2} \\ \hline 0 &  & A_\mu & \\ 1 & \lambda &  & \lambda' \\ 2 & & \phi &     \end{array} \;\;\;\; \;\;\;\;\;\;\;\;
\begin{array}{c|ccc}  R_{\mathcal{N}=2} \setminus I_3 & \frac{1}{2} & 0 & -\frac{1}{2} \\ \hline -1 &  & \psi & \\ 0 & Q &  & \widetilde{Q}^\dagger \\ 1 & & \widetilde{\psi} ^\dagger &     \end{array} \label{n2charges}
\ee  
Under these symmetries, the operators in a $T_N$ theory have charges
\be
 \begin{array}{c|ccc} & R_{\mathcal{N}=2} & &  I_3 \\ \hline
 u_k & 2k &  & 0 \\
 Q & 0  &  & \frac 1 2 (N-1) \\
 \widetilde Q & 0  &  & \frac 1 2 (N-1) \\
 \mu & 0&  & 1 \\
\end{array} \label{tn2charges}
\ee  

It is also useful to consider an $\mathcal{N}=1$ subalgebra of the $\mathcal{N}=2$ algebra, with the corresponding $\cN=1$ R-symmetry. Such a choice of subalgebra corresponds to a particular choice of Cartan generator $I_3$, in terms of which the $\cN=1$ R-symmetry is given by
\be
R_{\cN=1} = \frac 13 R_{\mathcal{N}=2} + \frac 43 I_3~. \label{R1app}
\ee
Although there are many other possible choices for an $\mathcal{N}=1$ R-symmetry, this choice is the unique one that has the properties of a superconformal $U(1)_R$ when the theory has $\cN=2$ SUSY. 

We can now compute the central charges of a $T_N$, as in \cite{Gaiotto:2009we}. They are given by
\be
\begin{split}
a_{T_N} &= \frac{3}{32} \left (3  \Tr R_{\cN=1}^3 - \Tr R_{\cN=1} \right ) = \frac{1}{24} \left ( 12 \Tr R_{\mathcal{N}=2} I_3^2 - \frac 12 \Tr R_{\mathcal{N}=2}^3 \right )~, \\[5pt]
 c_{T_N} &= \frac{1}{32} \left (9  \Tr R_{\cN=1}^3 - 5 \Tr R_{\cN=1} \right ) =  \frac{1}{12} \left ( 6 \Tr R_{\mathcal{N}=2} I_3^2 - \frac 12 \Tr R_{\mathcal{N}=2}^3 \right )~,
\end{split}
\ee
with the 't Hooft anomalies taking the following values,
\be\begin{split}
&\Tr R_{\mathcal{N}=2}^3 = \Tr R_{\mathcal{N}=2} = 2 + N - 3 N^2~, \\
&\Tr R_{\mathcal{N}=2} I_3^2 = \tfrac{1}{12} \left ( 6 - N - 9 N^2 + 4 N^3 \right)~.
\end{split}
\ee
The resulting values for the central charges are
\be
a_{T_N} = \frac{N^3}{6} - \frac{5 N^2}{16} - \frac{N}{16} + \frac{5}{24}~, \qquad\qquad c_{T_N} =  \frac{N^3}{6} - \frac{N^2}{4} - \frac{N}{12} + \frac{1}{6}~.
\ee

It is also crucial to know some 't Hooft anomalies for the non-Abelian currents. If $T^{a,b}$ are generators of an $SU(N)$ global symmetry, 
\be
\Tr R_{\cN=1}T^a T^b = - \frac{1}{3} N~.
\ee  
Moreover, in the $\cN=1$ language there is an additional $U(1)$ symmetry which commutes with the $\cN=1$ supercharges and is given by
\be
 J = R_{\mathcal{N}=2} - 2 I_3~. \label{Japp}
\ee
The anomaly contribution for $J$ is the same as that of $N$ fundamental hypermultiplets, 
\be
\Tr JT^a T^b = -N~.
\ee

\bibliographystyle{utphys}
\bibliography{longversion}{}
\end{document}